\documentclass[10pt]{article}
\pdfoutput=1
\usepackage{jheppub,amsmath,amssymb}
\usepackage{enumerate}
\usepackage{bm}
\usepackage{bbm}
\usepackage{enumitem}
\usepackage[usenames,dvipsnames]{xcolor}
\usepackage{amsmath}
\usepackage{amsfonts}
\usepackage{amssymb}
\usepackage{graphicx}   
\usepackage{hhline}
\usepackage{subfig}
\usepackage{hyperref}
\usepackage{color}
\usepackage{verbatim}
\usepackage{amsmath,amsthm,amssymb}
\usepackage{lscape}
\usepackage{float}
\usepackage{caption}
\usepackage{lscape}
\usepackage{breqn}
\usepackage{multicol}
\usepackage{graphics}
\usepackage{tikz,todonotes}
\usepackage[utf8]{inputenc}
\usepackage{autobreak}
\usepackage{verbatim}
\usepackage{tikz-feynman}
\usetikzlibrary{arrows,positioning,decorations.markings,decorations.pathmorphing,calc}
\pgfdeclarelayer{edgelayer}
\pgfdeclarelayer{nodelayer}
\usetikzlibrary{decorations.pathreplacing}
\pgfsetlayers{edgelayer,nodelayer,main}
\tikzset{none/.style={draw=none}}
\tikzset{new edge style 2/.style={black}}
\tikzset{new style 0/.style={black}}
\tikzset{rednode/.style={draw=none, scale=0.3pt,fill=red,circle, draw}}
\tikzset{redline/.style={line width=0.3mm,red}}
\tikzset{greyE/.style={line width=0.1mm,gray}}
\usepackage[utf8]{inputenc}
\usetikzlibrary{arrows,positioning,decorations.markings,decorations.pathmorphing,calc}

\definecolor{hyperref}{RGB}{026,028,087}

\newcommand{\beq}{\begin{equation}}
\newcommand{\eeq}{\end{equation}}
\newcommand{\bea}{\begin{eqnarray}}
\newcommand{\eea}{\end{eqnarray}}
\def\be{\begin{equation}}
\def\ee{\end{equation}}

\def\beq{\begin{equation}}
\def\eeq{\end{equation}}

\renewcommand{\L}{\mathcal L}

\def\be{\begin{equation}}
\def\ee{\end{equation}}
\def\ba{\begin{eqnarray}}
\def\ea{\end{eqnarray}}

\def\nn{\nonumber}

\usepackage[normalem]{ulem}

\def\ba{\begin{eqnarray}}
\def\ea{\end{eqnarray}}

\def\L{\mathcal{L}}

\def\stu{St\"uckelberg }

\def\({\left(}
\def\){\right)}

\def\ie{{\em {\it i.e.} }}

\allowdisplaybreaks
\begin{document}

\title{Kaluza-Klein from Colour-Kinematics Duality for Massive Fields}

\author[a]{Arshia Momeni,}
\author[a]{Justinas Rumbutis,}
\author[a,b]{Andrew J. Tolley}

\affiliation[a]{Theoretical Physics, Blackett Laboratory, Imperial College, London, SW7 2AZ, U.K.}
\affiliation[b]{CERCA, Department of Physics, Case Western Reserve University, 10900 Euclid Ave, Cleveland, OH 44106, USA}

\emailAdd{arshia.momeni17@imperial.ac.uk}
\emailAdd{j.rumbutis18@imperial.ac.uk}
\emailAdd{a.tolley@imperial.ac.uk}

\abstract{We consider a broad class of massive four dimensional effective theories describing an infinite tower of charged massive spin 1 states, interacting with massless spin 1 and spin 0. The spectrum is chosen to be the same as that appears in the Kaluza-Klein theory reduction of 5d Yang-Mills to ensure the absence of any spurious poles in a possible double copy formulation. The effective theories are characterized by multiple different couplings between different fields which are unconstrained by symmetries and low energy criteria. Remarkably, by demanding that the scattering amplitudes preserve colour-kinematics duality for different scattering processes, required for the existence of a massive double copy, we find that our 4d Lagrangian is fixed uniquely to the Kaluza-Klein compactification of 5d Yang-Mills theory together with its known double copy consistent higher derivative operators. }

\maketitle


\section{Introduction}
There are now multiple examples of the double copy relation between scattering amplitudes of two different theories that work at all orders \cite{Kawai:1985xq, Bern:2010ue, Cachazo:2014xea, Cheung:2017ems, Chen:2013fya}. There are many applications of the double copy paradigm, ranging from relations between classical solutions of two theories, also known as classical double copy, \cite{Saotome:2012vy, Monteiro:2014cda, Luna:2015paa, Luna:2016due, White:2016jzc, Cardoso:2016amd, Luna:2016hge, Goldberger:2017frp, Ridgway:2015fdl, De_Smet_2017, Bahjat_Abbas_2017, carrillogonzalez2017classical, Goldberger_2018, Li_2018, Lee_2018, Plefka_2019, Berman_2019, Kim:2019jwm, Goldberger:2019xef, Alawadhi:2019urr, Banerjee:2019saj,Huang:2019cja, Berman:2020xvs}, simplifying gravitational calculations \cite{Neill:2013wsa,Bern:2019crd,Shen:2018ebu,Bautista:2019tdr,Cheung:2020gyp}, to considerations of effective field theories \cite{Carrillo-Gonzalez:2019aao, Low:2019wuv, Carrasco:2019qwr, Carrasco:2019yyn,Low:2020ubn, Cheung:2020qxc,Rodina:2020jlw}. A full understanding of the applicability of the double copy paradigm to effective field theories is yet to be obtained, however it would be of significant phenomenological interest. Almost of all of these developments are well established for massless theories, while the double copy for massive theories is far less understood. There are some known examples of amplitudes in theories in which certain states are massive compatible with colour-kinematics duality \cite{Bern:2008qj} (which is necessary precursor for the double copy) for example, gauge theories with massive matter  \cite{Naculich:2014naa,Chiodaroli:2015rdg,Johansson:2015oia,Chiodaroli:2017ehv,Chiodaroli:2018dbu,bautista2019double, Bjerrum-Bohr:2019nws,Johansson:2019dnu, Plefka:2019wyg, bautista2019scattering, Carrasco:2020ywq}, topologically massive theories \cite{Moynihan:2020ejh}, but the general extent to which the double copy paradigm applies to theories with massive states is not known.\\

Recently, there have been attempts to directly develop a double copy for massive spin-1 theories in four dimensions, \cite{Momeni:2020vvr, Johnson:2020pny} for which the naive expectation is that the double copy theory is an interacting massive spin-2 theory, i.e. massive gravity \cite{deRham:2010ik,deRham:2010kj}. This possibility is strongly suggested by the fact that the decoupling limit of massive Yang-Mills is a non-linear sigma model, and the latter is known to admit an all orders double copy as a special Galileon \cite{Hinterbichler:2015pqa}. Intriguingly, the Galileon theories are known in turn to arise as decoupling limits of massive gravity theories \cite{deRham:2010ik,deRham:2010kj}, suggesting a possible uplift to a full gravitational theory. While it was found in \cite{Momeni:2020vvr, Johnson:2020pny} that the naive double copy recipe for massive Yang-Mills does match precisely a consistent interacting massive spin-2 theory at 3pt and 4pt level, it was noted in \cite{Johnson:2020pny} that higher points amplitudes contain spurious poles at unphysical locations preventing an interpretation of the resulting double copy amplitudes as a local field theory. In particular certain spectral conditions were derived in \cite{ Johnson:2020pny}, which we review in section~\ref{sec:spec}, which are necessary to be satisfied to avoid the spurious poles at higher order. It remains unclear whether these spurious poles can be removed by adding new irrelevant operators or new fields to the massive Yang-Mills action. For example, as suggested in \cite{Chiodaroli:2015rdg} the spurious poles could signal a presence of a new massive state that has to be included in the spectrum.\\

On the other hand, the spectral conditions are at least satisfied for one well known theory of interacting massive spin-1 states: Kaluza-Klein (KK) gauge theory. It is known that for 5 dimensional (5d) theories on 4d Minkowski $\times S^1$ which are compatible with colour-kinematics duality, the double copy procedure works after we compactify the theory to four dimensions (4d). The condition for the absence of spurious poles in 4d is essentially automatically satisfied by virtue of the kinematics and mass spectrum implied from 5d, together with the charge conservation inherited from 5d compactified translation invariance. Thus as an example, pure 5d Yang-Mills compactified on an $S^1$ automatically gives a 4d theory of interacting massive spin-1 states for which colour-kinematics duality is respected, and for which there is a known double copy. \\

In this work we aim to explore whether there are other consistent effective field theories (EFTs) for interacting massive spin-1 states, which still respect colour-kinematics duality, for which there is a possibility of developing a double-copy. Since the conditions necessary to remove the spurious poles are highly non-trivial, and it is not straightforward to satisfy them, we will rather fix the spectrum and charge assignments of our interacting effective field theory to be identical to that of 5d Yang-Mills compactified on an $S^1$. This allows us to avoid the automatic appearance of spurious poles, and still gives us considerable freedom in choosing local interactions. In particular, we do not require that our 4d theory respects the full compactified 5d gauge invariance, rather only that it preserves 4d gauge invariance, together with a global $U(1)$ symmetry which is the remnant of 5d translations. There are then a huge number of distinct operators which can be included in the EFT which are consistent with the symmetries and are unconstrained by other low energy criteria.\footnote{We do expect these couplings to be constrained by positivity bounds on the scattering of massive spin-1 states similar to those considered in \cite{deRham:2017zjm,deRham:2018qqo}.} Our goal then is to explore this huge space of effective field theories to establish whether there are other possible cases which admit colour-kinematics duality. \\

Remarkably we find that if we demand the naive extension of colour-kinematics (CK) duality for massive states proposed in \cite{Momeni:2020vvr, Johnson:2020pny} to an array of scattering amplitudes with different external states, the huge freedom in our EFT coupling constant space is reduced, and to the order we have calculated, namely up to 5pt amplitudes and order $1/\Lambda^4$ in an EFT expansion, the unique solution which allows for CK duality is the Kaluza-Klein compactification of the 5d theory 
\be
{\cal L}_{5d} = \left(\frac{-1}{4}\text{tr}(F^2)+\frac{G_{5d}}{\Lambda^2}\text{tr}(F^3)-\frac{9G_{5d}^{2}}{16\Lambda^4}\text{tr}([F,F]^2)\right)+ {\cal O}(\Lambda^{-6})
\ee
which in the uncompactified 5d limit is known to admit a double copy. In this sense, and to the order we have calculated, we are in effect able to `derive' Kaluza-Klein theory as the only consistent EFT which admits colour-kinematics duality for a given spectrum of states and charges. Whilst this result is very powerful, it is also disappointing in terms of limiting the search for other examples of interacting massive theories which may admit a double copy.  \\

Our results stop short of being a proof of the inevitability of Kaluza-Klein given our chosen spectrum of states as we have not considered all possible EFT operators that arise up to the calculated EFT order $1/\Lambda^4$, nor have we computed 6pt or higher amplitudes. At each order a multitude of new contact terms can be included in the EFT expansion. Nevertheless, it becomes quite readily apparent that as the order of amplitude is increased, given the increasing number of constraints required in order to satisfy colour-kinematics, it becomes increasingly hard to find any remaining freedom beyond that from the compactification of the known CK compliant higher derivative operators. \\

We begin in section~\ref{sec:spec} with a review of the conditions necessary to remove spurious poles and a brief derivation of the BCJ relations for massive theories, with some mild assumption on the exchanged particles. In section~\ref{sec:action} we specify the large class of effective field theories we consider and begin the process of putting constraints on the multiple different coupling constants by demanding that the 4pt amplitudes respect the BCJ relations. Whilst this process leaves some freedom, this remaining freedom is removed by considering BCJ relations for the 5pt function as we do in section~\ref{sec:5pt}. In section~\ref{nonsym} we briefly consider some alternative possible EFT operators before concluding in section~\ref{conclusion}.

\section{Spectral conditions and BCJ relations}\label{sec:spec}

In this section we will review why constructing a theory with massive spin 1 states that respects colour-kinematics duality is difficult due to the potential emergence of spurious poles and the conditions necessary to remove them.  To double copy an $n$-point tree level gauge theory amplitude, ${\cal A}_n$, first we need to write it in the following form:
\begin{equation}\label{eq:An}
{\cal A}_n=\sum_{i}\frac{c_i n_i}{\prod_{\alpha_{i}}D_{\alpha_i}}=c^T D^{-1}n, 
\end{equation}
where $c_i$ are the colour factors, $n_i$ are the kinematic numerators and $D_{\alpha_i}$ are the inverse propagators, The second equality is the first one written in a condensed matrix form, so that now $c$ is a column vector of colour factors, $D$ is diagonal matrix whose elements are the products of inverse propagators, and $n$ is the column vector of kinematic numerators. Our starting assumption is that this representation for the amplitude for massive states in \eqref{eq:An} must satisfy the colour-kinematics duality \cite{Bern:2008qj} known to be satisfied for massless theories. That is, whenever three of the colour factors, $c_i$, $c_j$ and $c_k$ are related by a Jacobi identity, such as $c_i+c_j+c_k=0$, the corresponding kinematic factors must obey the same relation {\it i.e.} $n_i+n_j+n_k=0$. In matrix form this can be written as
\be
\label{eq:Mc}
Mc=0\rightarrow Mn=0 \, ,
\ee
for some matrix $M$ with entries $\pm1$. In  \cite{Momeni:2020vvr, Johnson:2020pny}  it was proposed to construct the double copy gravitational theory with amplitudes ${\cal M}_n$ in direct analogy with the massless double copy via
\begin{equation}\label{eq:Mn}
{\cal M}_n\propto\sum_{i}\frac{n_i \tilde n_i}{\prod_{\alpha_{i}}D_{\alpha_i}} = n^T D^{-1} \tilde n \, ,    
\end{equation}
so that the only difference between massless and massive is the replacement of massless propagators with massive ones. If this amplitude comes from a local action the location of its poles must correspond to the exchange of the masses in the spectrum of this gravitational theory. That can be achieved if the kinematic numerators, $n_i$, do not have unphysical (spurious) poles. It is obvious that such unphysical poles do not appear in the kinematic factor $n_i$ if they are directly calculated from Feynman diagrams of a local Lagrangian. However, such $n_i$ may not automatically satisfy Jacobi relations $Mn=0$ and it is in the attempted resolution of this that spurious poles may arise.\\

Let us assume that the calculated $n_i$ do not initially satisfy $Mn=0$. If this is the case then we can shift their values as
\be
n\rightarrow n+\Delta n,
\ee
such that the amplitude in Eq. \eqref{eq:An} is unchanged, which can be achieved by setting
\be
\label{eq:deltan}
\Delta n=D M^{T}v,
\ee
where $v$ is some as yet undetermined column vector. 
Explicitly to confirm that the amplitude is unchanged by this shift
\be
\Delta {\cal A}_n = c^T D^{-1} \Delta n = c^T M^{T}v= (Mc)^T v =0 \, ,
\ee
by virtue of the colour Jacobi identity $Mc=0$. In order to satisfy CK duality the shifted $n$ must obey the following equation
\be
M(n+\Delta n)=0,
\ee
which combined with Eq. \eqref{eq:deltan} gives
\be\label{eq:cond v}
MDM^{T}v=-Mn.
\ee
The number of non-zero rows of $M$ will be equal to the number of Jacobi identities, $N_j$. Therefore, the symmetric matrix $MDM^{T}$ will be block diagonal containing an $N_j \times N_j$ symmetric block matrix $A$ with all other elements equal to zero. Similarly $Mn$ will have at most $N_j$ non-zero elements. Therefore, to find the shifts $\Delta n$, we need to find $v$, for which we need to invert the $N_j \times N_j$ matrix $A$.  \\

In a generic theory, $A$ will be invertible and the inverse of $A$ will have elements with poles at locations which are some complicated expressions of masses in the theory giving spurious poles in the shifted $n$. Specifically the location of the spurious poles will be determined by $\text{det} A=0$ and since $\text{det} A$ is in general a complicated function of kinematic invariants and masses, the poles will typically be uncorrelated with those demanded by locality and unitarity. Hence while it is possible to construct kinematic factors which respect CK duality, the resulting gravitational theory defined by \eqref{eq:Mn} cannot be interpreted as a local field theory. This is for example the situation that arises for the proposed double copy of massive Yang-Mills as discussed in \cite{Momeni:2020vvr, Johnson:2020pny} at the level of the 5pt amplitude.\\

Once this is recognized, it becomes apparent how to `solve' the problem by tuning the theory such that $A$ is singular, i.e. has reduced rank\footnote{The spectral conditions derived in \cite{Johnson:2020pny} come from requiring the KLT matrix of bi-adjoint scalar amplitudes to be singular/have reduced rank but are equivalent to $\text{det}A=0$ in our language.}. In this case there is no unique solution for $v$, but if $Mn$ further obeys certain conditions (it should be in the image of $A$) Eq. \eqref{eq:cond v} can still be solved. These conditions on $n$ give precisely the BCJ relations. When $A$ is singular, the rank of $A$ is necessarily less than $N_j$ and by reducing it by the right number of conditions the expression for $v$ can be determined, without spurious poles provided there are no spurious zeros in the determinant of the reduced rank $A$. This is of course exactly what happens for pure Yang-Mills theory for which the double copy procedure is well established for all tree-level amplitudes. In general to avoid spurious poles not only $A$ has to be singular but its rank should be equal to that of pure Yang-Mills to ensure the correct number of BCJ relations. \\

If we restrict to the case of a single exchanged mass per channel and fields in the adjoint representation, we can satisfy these conditions if the elements of $D$ obey the same algebraic relations as that for pure Yang-Mills theory. For example in the case of 4pt pure Yang-Mills we have a single Jacobi identity, $c_s+c_t+c_u=0$ and $D=\text{diag}\{s,t,u\}$ which obeys in the massless case $\text{Tr}D=s+t+u=0$. In this case we only have one Jacobi identity at 4pt so $M$ has only one non-zero row, $\{1,1,1\}$. Now if we have a massive theory and we scatter $m_1,m_2,m_3,m_4$ with $D=\text{diag}\{s-m_{12}^2,t-m_{13}^2,u-m_{14}^2\}$, one may easily show that the only nonzero entry in $MDM^{T}$ is $\text{Tr}D$, hence the $1 \times 1 $ matrix $A$ is simply $\text{Tr}D$. In the massless theory $\text{Tr}D$ is identically zero and so the rank of $A$ is zero. Enforcing the requirement that the rank is also reduced to zero in the massive theory leads to
\be\label{spectral1}
s+t+u-m_{12}^2-m_{13}^2-m_{14}^{2}=m_1^2+m_2^2+m_3^2+m_4^2-m_{12}^2-m_{13}^2-m_{14}^{2}=0,
\ee
which is the 4pt spectral condition proposed in \cite{Johnson:2020pny}. Since the rank of $A$ has been reduced by 1, we obtain one BCJ relation which is obviously the one non-zero element of Eq. \eqref{eq:cond v} $Mn=0$, \ie
\be
n_s+n_t+n_u=0,
\ee
for the $n$'s calculated directly from Feynman diagrams. This is of course the standard 4pt BCJ relation. There is a BCFW recursion proof of BCJ relations \cite{Feng:2010my,Jia:2010nz,Chen:2011jxa} so in theories for which BCFW recursion works, like pure YM, the lower point BCJ relations (and spectral conditions) imply the higher point ones.
We should note that from 4pt considerations alone, it is not necessary to impose \eqref{spectral1} since in this case $\det A$ is a constant and so no spurious poles arise from the 4pt function. This is exactly why we are able to develop a double copy for massive Yang-Mills up to 4pt order without any problems as shown in \cite{Momeni:2020vvr, Johnson:2020pny}. This situation does not persist however at higher order. \\

At five points for a theory with all of the fields in the adjoint representation we have 15 colour factors given by
\begin{eqnarray}
&& 
c_{1\phantom{0}} \equiv f^{a_1 a_2 b}f^{b a_3 c}f^{c a_4 a_5}\,, \hskip 0.8cm  
c_{2\phantom{1}} \equiv f^{a_2 a_3 b}f^{b a_4 c}f^{c a_5 a_1}\,, \hskip 0.8cm  
c_{3\phantom{1}} \equiv f^{a_3 a_4 b}f^{b a_5 c}f^{c a_1 a_2}\,, \nn \\&&
c_{4\phantom{1}} \equiv f^{a_4 a_5 b}f^{b a_1 c}f^{c a_2 a_3}\,, \hskip 0.8cm  
c_{5\phantom{1}} \equiv f^{a_5 a_1 b}f^{b a_2 c}f^{c a_3 a_4}\,, \hskip 0.8cm  
c_{6\phantom{1}} \equiv f^{a_1 a_4 b}f^{b a_3 c}f^{c a_2 a_5}\,, \nn \\&& 
c_{7\phantom{1}} \equiv f^{a_3 a_2 b}f^{b a_5 c}f^{c a_1 a_4}\,, \hskip 0.8cm  
c_{8\phantom{1}} \equiv f^{a_2 a_5 b}f^{b a_1 c}f^{c a_4 a_3}\,, \hskip 0.8cm  
c_{9\phantom{1}} \equiv f^{a_1 a_3 b}f^{b a_4 c}f^{c a_2 a_5}\,, \nn \\&&
c_{10} \equiv f^{a_4 a_2 b}f^{b a_5 c}f^{c a_1 a_3}\,, \hskip 0.8cm  
c_{11} \equiv f^{a_5 a_1 b}f^{b a_3 c}f^{c a_4 a_2}\,, \hskip 0.8cm  
c_{12} \equiv f^{a_1 a_2 b}f^{b a_4 c}f^{c a_3 a_5}\,, \nn \\\ &&
c_{13} \equiv f^{a_3 a_5 b}f^{b a_1 c}f^{c a_2 a_4}\,, \hskip 0.8cm  
c_{14} \equiv f^{a_1 a_4 b}f^{b a_2 c}f^{c a_3 a_5}\,, \hskip 0.8cm  
c_{15} \equiv f^{a_1 a_3 b}f^{b a_2 c}f^{c a_4 a_5}\,. \hskip 1.5 cm 
\label{FivePointColor}
\end{eqnarray} 
There are 9 independent Jacobi identities that can be written in the form \eqref{eq:Mc} with the matrix $M$ given by
\begin{eqnarray}
M=\left(
\begin{array}{ccccccccccccccc}
 0 & 0 & 1 & 0 & -1 & 0 & 0 & 1 & 0 & 0 & 0 & 0 & 0 & 0 & 0 \\
 -1 & 0 & 1 & 0 & 0 & 0 & 0 & 0 & 0 & 0 & 0 & 1 & 0 & 0 & 0 \\
 -1 & 0 & 0 & 1 & 0 & 0 & 0 & 0 & 0 & 0 & 0 & 0 & 0 & 0 & 1 \\
 0 & -1 & 0 & 1 & 0 & 0 & 1 & 0 & 0 & 0 & 0 & 0 & 0 & 0 & 0 \\
 0 & -1 & 0 & 0 & 1 & 0 & 0 & 0 & 0 & 0 & 1 & 0 & 0 & 0 & 0 \\
 0 & 0 & 0 & 0 & 0 & -1 & 1 & 0 & 0 & 0 & 0 & 0 & 0 & 1 & 0 \\
 0 & 0 & 0 & 0 & 0 & -1 & 0 & 1 & 1 & 0 & 0 & 0 & 0 & 0 & 0 \\
 0 & 0 & 0 & 0 & 0 & 0 & 0 & 0 & -1 & 1 & 0 & 0 & 0 & 0 & 1 \\
 0 & 0 & 0 & 0 & 0 & 0 & 0 & 0 & 0 & 1 & -1 & 0 & 1 & 0 & 0 \\
 0 & 0 & 0 & 0 & 0 & 0 & 0 & 0 & 0 & 0 & 0 & 0 & 0 & 0 & 0 \\
 0 & 0 & 0 & 0 & 0 & 0 & 0 & 0 & 0 & 0 & 0 & 0 & 0 & 0 & 0 \\
 0 & 0 & 0 & 0 & 0 & 0 & 0 & 0 & 0 & 0 & 0 & 0 & 0 & 0 & 0 \\
 0 & 0 & 0 & 0 & 0 & 0 & 0 & 0 & 0 & 0 & 0 & 0 & 0 & 0 & 0 \\
 0 & 0 & 0 & 0 & 0 & 0 & 0 & 0 & 0 & 0 & 0 & 0 & 0 & 0 & 0 \\
 0 & 0 & 0 & 0 & 0 & 0 & 0 & 0 & 0 & 0 & 0 & 0 & 0 & 0 & 0 \\
\end{array}
\right).
\end{eqnarray}
Assuming again that there is just a single mass exchange for each colour factor the $D$ matrix is given by
\begin{align}
 D=&\text{diag}\{ D_{12} D_{45}, D_{15} D_{23}, D_{12} D_{34}, D_{23} D_{45}, D_{15} D_{34}, D_{14} D_{25}, D_{14} D_{23}, D_{25}
D_{34},  \nn \\&D_{13} D_{25}, D_{13} D_{24}, D_{15} D_{24}, D_{12} D_{35}, D_{24} D_{35}, D_{14} D_{35}, D_{13} D_{45}\},
\end{align}
where $D_{ij}=-(p_i+p_j)^2-m_{ij}^2=s_{ij}-m_{ij}^2$.
Now we require the non-zero $9 \times 9$ block matrix in \eqref{eq:cond v}, $A$, to be of rank 5, such that there are 4 BCJ relations between the $n$'s just like in pure Yang-Mills case. 
By imposing the five 5pt spectral conditions \cite{Johnson:2020pny}
\ba\label{spectral2}
&&m_{15}^2 = 2 m_1^2 - m_{12}^2 - m_{13}^2- m_{14}^2+m_2^2+m_3^2+m_4^2+m_5^2 \nn  \\
&& m_{25}^2=m_1^2-m_{12}^2+2 m_2^2-m_{23}^2-m_{24}^2+m_3^2+m_4^2+m_5^2 \nn \\
&& m_{34}^2=2m_1^2-m_{12}^2-m_{13}^2-m_{14}^2+2m_2^2-m_{23}^2-m_{24}^2+2m_3^2+2m_4^2+m_5^2 \nn \\
&& m_{35}^2=-m_1^2+m_{12}^2+m_{14}^2-m_2^2+m_{24}^2-m_4^2  \nn \\
&& m_{45}^2=-m_1^2+m_{12}^2+m_{13}^2-m_2^2+m_{23}^2-m_3^2 \, ,
\ea
one can show that the symmetric matrix $A$, which is given explicity in appendix~\ref{appendixBCJ}, indeed has rank 5. The conditions \eqref{spectral2} are justified by demanding that that 4pt spectral condition is satisfied on every 4-point amplitude that arises in a factorization channel for the 5pt amplitude such as shown in Fig.~\ref{fig: facto}. Since $A$ has rank $5$ it admits four null-eigenvectors $u_\alpha$ (see \eqref{nullspace}).  We promote the number of components of the null-eigenvector $u_\alpha$  to 15 by adding six zeros , $U^T_\alpha=(u_\alpha^T,0,0,0,0,0,0)$,  then from \eqref{eq:cond v} we have for each null eigenvector
\be\label{BCJrelations}
U_\alpha^T M n =0 \, .
\ee
The 5pt BCJ relations written in \cite{Johnson:2020pny} in terms of kinematic factors before applying the spectral conditions are,
\begin{equation}
\label{eq:bcj1}
\begin{split} 
   &D_{12}^2 (-(D_{15}D_{34} n_{4}+D_{23}D_{45} n_{5}+D_{34}D_{45} n_{2}))+D_{12} (D_{15} (D_{23} (D_{14}
   n_{8}+D_{34} (-n_{1}+n_{13}+n_{6})\\&-D_{45} n_{3})-D_{24}D_{34} n_{4}+D_{25}D_{34} n_{7})+D_{23} (D_{14}
  D_{25} n_{5}-D_{24} n_{5} (D_{35}+D_{45})\\&+D_{34}D_{35} n_{11})+D_{34} n_{2} (D_{14}D_{25}-D_{24}
  D_{45}))-D_{15}D_{23} D_{24} (D_{34} (n_{1}-n_{12})+n_{3} (D_{35}+D_{45}))=0,
\end{split}  
\end{equation}
  \begin{equation}
  \label{eq:bcj2}
\begin{split}
  &D_{14} (D_{15} D_{34} (D_{12} n_{4}-D_{25} n_{7})-D_{23} (-D_{12} D_{45} n_{5}+D_{24} D_{25} n_{5}+D_{34} D_{35}
   n_{11})+D_{12} D_{34} D_{45} n_{2}\\&+D_{15} D_{23} (-D_{24} n_{8}+D_{34} (n_{1}-n_{13}-n_{6})+D_{45}
   n_{3})-D_{24} D_{34} n_{2} (D_{25}+D_{35}))\\&+D_{14}^2 (-(D_{15} D_{23} n_{8}+D_{23} D_{25} n_{5}+D_{25} D_{34}
   n_{2}))+D_{15} D_{24} D_{34} (D_{23} (n_{14}-n_{6})-n_{7} (D_{25}+D_{35}))=0,
\end{split}  
\end{equation}
\begin{equation}
\label{eq:bcj3}
\begin{split}
   &-D_{15} (D_{12} D_{34} n_{4} (D_{24}+D_{45})+D_{13} D_{24} D_{34} n_{4}+D_{23} D_{45} (-D_{14} n_{8}+D_{34}
   (n_{1}+n_{10}-n_{6})+D_{45} n_{3})\\&+D_{23} D_{24} (D_{34} (n_{1}-n_{15})+D_{45} n_{3})-D_{25} D_{34} D_{45}
   n_{7})-D_{45} ((D_{23} n_{5}+D_{34} n_{2}) (D_{12} (D_{24}+D_{45})-D_{14} D_{25})\\&+D_{13} D_{34} (D_{23}
   n_{11}+D_{24} n_{2}))=0,
\end{split}  
\end{equation}
\begin{equation}
\label{eq:bcj4}
\begin{split}
        &D_{15} (D_{12} D_{25} D_{34} n_{4}-D_{23} D_{24} (n_{8} (D_{13}+D_{14})+D_{34} (n_{6}-n_{9}))+D_{23} D_{25}
   (-D_{14} n_{8}+D_{34} (n_{1}+n_{10}-n_{6})\\&+D_{45} n_{3})-D_{25} D_{34} n_{7} (D_{24}+D_{25}))-D_{25} ((D_{23}
   n_{5}+D_{34} n_{2}) (D_{14} (D_{24}+D_{25})-D_{12} D_{45})\\&+D_{13} D_{23} (D_{24} n_{5}-D_{34} n_{11}))=0,
\end{split}  
\end{equation}
once the spectral condition is applied, linear combinations of $U^T_\alpha$ in \eqref{BCJrelations} can reproduce the four BCJ relations (see \eqref{linearcomb}). These give the four BCJ relations expressed in terms of kinematic numerators rather than partial amplitudes. In practise, these are the same algebraic relations that arise in the massless case,  with the replacement $s_{ij} \rightarrow D_{ij}$ in \cite{Bern:2008qj}. In other words the massive double copy formalism developed in \cite{Momeni:2020vvr, Johnson:2020pny} assumes that all algebraic relations between partial amplitudes and propagators that are true in massless theories, compatible with the colour-kinematics duality, should still hold for massive ones. \\

This general procedure straightforwardly generalizes to higher $n$-pt functions. At each order, there will be a set of spectral conditions necessary to reduce the rank of $A$ by the number of required BCJ relations at that order. The latter are then determined by the null eigenvectors of the reduced rank matrix from \eqref{BCJrelations}. Ultimately the key to the existence of a local double copy is that the reduced rank matrix does admit an inverse without spurious poles. It is not obvious to see why this is from our formal BCJ language argument, however this can be elucidated using the related KLT formalism used in \cite{Johnson:2020pny}. There the role of our BCJ matrix $A$ is played by the KLT matrix of bi-adjoint scalar amplitudes, and the spectral conditions may be similarly derived from the requirement that the rank of the matrix of bi-adjoint scalar amplitudes is reduced to the `minimal' value.\\

In KLT language the BCJ relations make the double copy answer independent of the choice of which partial gauge theory amplitudes we use as an input, while in this BCJ language this is reflected by generalised gauge freedom, \ie the freedom to choose $\Delta n$ in \eqref{eq:deltan} when it is not uniquely fixed by \eqref{eq:cond v} (that is when $A$ is singular).\\

As mentioned in the introduction, known example of theories with massive states that will automatically satisfy the above spectral conditions \eqref{spectral1} and \eqref{spectral2} are Kaluza-Klein theories obtained by compatifying 5d massless theories obeying the colour-kinematics duality. The reason for this is simply that the massive propagator $D_{ij}$ can always be interpreted as a massless propagator in 5 dimensions for a specific choice of 5d momenta. Since the 5d kinematic factors do satisfy the BCJ relations, then evaluating them on the appropriate compactified 5d momenta, they automatically descend into 4d. Furthermore the spectral conditions \eqref{spectral1} and \eqref{spectral2} follow automatically from 5d momentum conservation which in 4d terms can be interpreted as charge conservation for a global $U(1)$ charge. It remains unclear whether there are other nontrivial solutions of these spectral conditions other than that given by Kaluza-Klein theories. Given this, in what follows we shall follow the opposite approach. We shall construct effective theories with the same spectrum and global $U(1)$ charge conservation properties as Kaluza-Klein, so that the spectral conditions \eqref{spectral1} and \eqref{spectral2} are automatically satisfied. We then ask what freedom there exists in the form of their interactions, which will necessarily alter the kinematic factors $n_i$, such that they still satisfy the BCJ relations \eqref{eq:bcj1}--\eqref{eq:bcj4}. According to our procedure, as long as the latter are satisfied, it is possible to solve for $v$, hence for $\Delta n$, and hence determine new kinematic factors $n+\Delta n$ which do respect CK duality, from which a double copy theory may be constructed given via the standard prescription:
\begin{equation}\label{eq:Mn}
{\cal M}_n\propto (n+\Delta n)^T D^{-1} (\tilde n+\Delta \tilde n) \, .
\end{equation}

\section{KK inspired action}\label{sec:action}

We consider a 4d effective field theory of interacting massless scalar and massive and massless vectors fields with the same spectrum as Kaluza-Klein theory. In particular, the massive states will be charged under a global $U(1)$ symmetry, the remnant of now forgotten 5d translations, and the mass will be proportional to the charge, with the infinite spectrum of charges integer spaced. The massive vectors will further transform in the adjoint for some gauge group $G$. We shall express the action for the massive states in 4d unitary gauge\footnote{This is unitary gauge for the gauge symmetries which are broken by the mass for the spin-1 states and has nothing to do with the unbroken 4d gauge symmetry $G$.}, \ie we will not introduce any \stu fields which arise naturally from compactification from the higher dimensional gauge symmetry \cite{Bonifacio:2019ioc}, meaning that the quadratic part of the action for the massive states is a complex gauged Proca theory. The remaining 4d gauge symmetry, the gauge freedom of the 4d massless gluon, is however made manifest. \\

In the EFT context, there is still a huge freedom in the choice of interactions between the states, even given the assumed 4d gauge symmetry and global $U(1)$ symmetry. In order to make calculational progress we will restrict to what remains a very large set of possible interactions. This set is chosen by the requirement that all the terms in the 4d action do indeed arise from compactification of 5d pure Yang-Mills together with additional $\frac{1}{\Lambda^2}\text{tr}(F^3)$ and $\frac{-9}{16\Lambda^4}\text{tr}([F_{\mu\nu},F_{\alpha\beta}][F^{\mu\nu},F^{\alpha\beta}])$ operators. We chose these higher order operators because it was shown in \cite{broedel2012color} that they are compatible with colour-kinematics duality. Crucially though we allow each of the coefficients of the various invariant terms in the action to be arbitrary, and in principle to be different for different interacting massive states. In doing so, the 4d action loses any further remnant information of the underlying 5d gauge symmetry. Thus while KK theory lies as a special point in our chosen class of EFTs, the class as a whole is still huge. 

\begin{table}[t]
    \centering
    \begin{tabular}{|c|c|c|c|c|c|}
    \hline
         $ $ & coefficient & Interactions  & $ $ & coefficient & Interactions \\
         \hline\hline
         $\L_{AAA}$ & $g_{ijk}$ & $ DAAA$ &   $\L^{F^4}_{AAAA1}$ & $c_{ijkl}$ & $(DA)^4$ \\
         \hline
         $\L_{AA\phi}$& $g'_{ijs}$ & $AA\phi$ & $\L^{F^4}_{AAAA2}$ & $C_{ijkl}$  & $m^2(DA)^2A^2$ \\
         \hline
         $\L_{AAAA}$ & $g_{ijkl}$ & $AAAA$ &   $\L^{F^4}_{AAA\phi1}$ & $c_{ijks}$ & $m(DA)^2A\phi$\\ 
         \hline
         $\L_{AA\phi\phi}$ & $g_{ijss}$ & $ AA\phi\phi$ & $\L^{F^4}_{AAA\phi2}$ & $C_{ijks}$ & $m^3A^3D\phi$ \\
         \hline
         $\L_{AAA^{0}}$ & $g_{i}$ & $AAF^{0}$ & $\L^{F^4}_{AA\phi\phi1}$ & $c_{ijss}$ & $(DA)^2(D\phi)^2$ \\
         \hline
         $\L^{F^3}_{AAA^{0}}$ & $G_{i}$ & $DADAF^{0}$ & $ \L^{F^4}_{AA\phi\phi2}$ & $c^{(2)}_{ijss}$  & $(mA)^2(D\phi)^2$ \\
         \hline
         $\L^{F^3}_{AAA1}$ & $G_{ijk}$ & $ (DA)^3$ & $ \L^{F^4}_{AA\phi\phi3}$ & $c^{(3)}_{ijss}$  & $(mA)^2(D\phi)^2$ \\
         \hline
         $\L^{F^3}_{AAA2}$ & $\hat{G}_{ijk}$& $m_im_jDAAA$ & $ \L^{F^4}_{AA\phi\phi3}$ & $c^{(3)}_{ijss}$  & $(mA)^2(D\phi)^2$ \\
         \hline
         $\L^{F^3}_{AA\phi}$ & $G'_{ijs}$& $AD\phi DA$ & $ \L^{F^4}_{\phi\phi\phi\phi}$ & $c_{\phi 4}$ & $(D\phi)^4$ \\
         \hline
         $\L^{F^3}_{A\phi\phi}$ & $G_{0ss}$& $D\phi D\phi F^{0}$ & $\L^{F^4}_{AAAAA1}$ & $c_{ijklm}$ & $(DA)^3A^2$ \\
         \hline
         $\L^{F^3}_{AAAA1}$ & $G_{ijkl}$ & $DADAAA$ & $\L^{F^4}_{AAAAA2}$ & $C_{ijklm}$ & $m^2(A)^4DA$ \\
         \hline
         $\L^{F^3}_{AAAA2}$ & $\hat{G}_{ijkl}$& $m_im_jAAAA$ & $\L^{F^4}_{\phi AAAA1}$ & $c_{ijkls}$ & $m(A)^3DAD\phi$\\
         \hline
         $\L^{F^3}_{AA\phi\phi 1}$ & $G_{ijss}$& $AAD\phi D\phi$ & $\L^{F^4}_{\phi AAAA2}$ & $C_{ijklm}$ & $m(DA)^2(A)^2\phi$ \\
         \hline
         $\L^{F^3}_{AA\phi\phi 2}$ & $\hat{G}_{ijss}$ & $A\phi DAD\phi$ & $\L^{F^4}_{\phi AAAA3}$ & $\hat{C}_{ijkls}$ & $(m)^3(A)^4\phi$\\

         \hline
         $\L^{F^3}_{AAA\phi 1}$ & $\hat{G}_{ijks}$ & $m AAAD\phi$  & $\L^{F^4}_{\phi \phi AAA1}$ & $c_{ijkss}$ & $DA(A)^2(D\phi)^2$ \\
         \hline
        $\L^{F^3}_{AAA\phi 2}$ & $G_{ijks}$ & $m AADA\phi$ & $\L^{F^4}_{\phi \phi AAA2}$ & $C_{ijkss}$ & $A(DA)^2(D\phi)^2$\\
        \hline
        $\L^{F^3}_{AAAAA}$ & $G_{ijklm}$  &$DAAAAA$ & $\L^{F^4}_{\phi \phi AAA3}$ & $\hat{C}^{(3)}_{ijkss}$ & $m^2(A)^3D\phi\phi$\\
        \hline
        $\L^{F^3}_{\phi AAAA}$ & $G_{ijkls}$ & $mA\phi AAA$ & $\L^{F^4}_{\phi \phi AAA4}$ & $\hat{C}^{(4)}_{ijkss}$ & $m^2(A)^3D\phi\phi$\\
        \hline
        $\L^{F^3}_{\phi \phi AAA1}$ & $G_{ijkss}$ & $D\phi \phi AAA$ & $\L^{F^4}_{\phi \phi AAA5}$ & $\hat{C}^{(5)}_{ijkss}$ & $m^2(A)^3D\phi\phi$\\
        \hline
        $\L^{F^3}_{\phi \phi AAA2}$ & $\hat{G}_{ijkss}$ & $DA\phi A\phi A$ & $\L^{F^4}_{\phi \phi \phi AA1}$ & $c_{ijsss}$ & $mA^2D\phi D\phi \phi$ \\
        \hline
$\L^{F^4}_{\phi \phi \phi AA2}$ & $c^{(2)}_{ijsss}$ & $mA^2(D\phi)^2\phi$ & $\L^{F^4}_{\phi \phi \phi AA3}$ & $c^{(3)}_{ijsss}$ & $mA^2(D\phi)^2\phi$ \\
\hline    
    \end{tabular}
    \caption{Coefficients of the interactions.}
    \label{tab:my_label}
\end{table}
Specifically then, we have an action with an uncharged massless scalar field $\phi$ and multiple charged massive vector fields $A^i$ transforming as matter fields in adjoint representation of some non-Abelian group $G$ for which $A^0$ is the gauge connection. To order $1/\Lambda^4$ the Lagrangian is given as:
\be\label{eq:lagr mYM}
\L=\L^{F^2}+\L^{F^3}+\L^{F^4},
\ee
where $\L^{F^2}$, $\L^{F^3}$ and $\L^{F^4}$ contain the operators appearing in the compactification of $\text{tr}(F^2)$, $\text{tr}(F^3)$ and $\text{tr}([F_{\mu\nu},F_{\alpha\beta}][F^{\mu\nu},F^{\alpha\beta}])$ respectively that contribute to up to 5pt amplitudes. They are as follows:
\ba
    \L^{F^2}&=\text{tr}\left(-\frac{1}{2}D_{\mu}\phi D^{\mu}\phi-\frac{1}{4}F^{0}_{\mu\nu}F^{0\mu\nu}-\frac{1}{2}\sum_{i\in \mathbb{Z}_{\ne 0}} \frac{1}{2}|D_{\mu}A^{i}_{\nu}-D_{\nu}A^{i}_{\mu}|^{2}-2g_i A^{i\mu}A^{-i\nu}F^{0}_{\mu\nu}+m^{2}_{i}|A^{i}_{\mu}|^{2}\right) \nn \\&+\L_{AAA}+\L_{AA\phi}+\L_{AA\phi\phi}+\L_{AAAA} \, ,
  \ea  
    \ba
     \L^{F^3}=&&\frac{1}{\Lambda^2}\text{tr}\left(G F^{0}_{\mu\nu}F^{0\nu\rho}F^{0\mu}_{\rho}+\sum_{i\in \mathbb{Z}_{\ne 0}}\bigg(3G_{i}D_{\mu}A^{i\nu}D_{\nu}A^{-i\rho}F_{0\rho}^{\mu}\bigg)\right) \nn \\
    &&+\L^{F^3}_{AAA1}+\L^{F^3}_{AAA2}+\L^{F^3}_{AA\phi}+\L^{F^3}_{A\phi\phi}+\L^{F^3}_{AAAA1}+\L^{F^3}_{AAAA2}+\L^{F^3}_{AA\phi\phi 1}+\L^{F^3}_{AA\phi\phi 2}+\L^{F^3}_{AAA\phi 1}+\L^{F^3}_{AAA\phi 2} \nn\\
    &&+\L^{F^3}_{AAAAA}+\L^{F^3}_{\phi AAAA}+\L^{F^3}_{\phi \phi AAA1}+\L^{F^3}_{\phi \phi AAA2} \, ,
    \ea
    \ba
    \L^{F^4}&&=\L^{F^4}_{AAAA1}+\L^{F^4}_{AAAA2}+\L^{F^4}_{AAA\phi1}+\L^{F^4}_{AAA\phi2}+\L^{F^4}_{AA\phi\phi1}+\L^{F^4}_{AA\phi\phi2}+\L^{F^4}_{AA\phi\phi3}+\L^{F^4}_{\phi\phi\phi\phi} \nn \\
    &&+\L^{F^4}_{AAAAA1}+\L^{F^4}_{AAAAA2}+\L^{F^4}_{\phi AAAA1}+\L^{F^4}_{\phi AAAA2}+\L^{F^4}_{\phi AAAA3}+\L^{F^4}_{\phi \phi AAA1}+\L^{F^4}_{\phi \phi AAA2}+\L^{F^4}_{\phi \phi AAA3} \nn \\
    &&+\L^{F^4}_{\phi \phi \phi AA1}+\L^{F^4}_{\phi \phi \phi AA2},
\ea
where,
\be
    D_{\mu}=\partial_{\mu}+i g A^{0}_{\mu} \, , \quad \quad     F^{0a}_{\mu\nu}=\partial_{\mu}A^{0a}_{\nu}-\partial_{\nu}A^{0a}_{\mu}+\frac{g}{\sqrt{2}}f^{abc}A^{0b}_{\mu}A^{0c}_{\nu} \, ,
\ee
and schematically, the interacting terms with the relevant coefficients are in table \ref{tab:my_label}. The exact terms are given in the appendix~\ref{interactingterms}. \\

We write couplings only of distinct operators coming from the compactification of the 5d Yang-Mills. The couplings of identical operators coming from different 5d terms are combined. For example, we get the interactions of the form
\be\label{Lf2 and Lf4}
f^{abe}f^{cde}\sum_{i,j,k,l \in \mathbb{Z}_{\ne 0}}\; \left(A^{ia}_{[\mu}A^{jb}_{\nu]}A^{kc[\mu}A^{ld\nu]}\right),
\ee
from the compactification of $\L^{F^2}$ as well as $\L^{F^4}$, which has $\frac{1}{\Lambda^4}$. We combine the couplings into a single one, $g_{ijkl}$. Hence, the coupling $g_{ijkl}$ is dependent on the scale $\Lambda$. Note that in KK theory $g_{ijkl}=g^2+\frac{18m_{1}m_{2}m_{3}m_{4}}{\Lambda^4}G^2$.\\

Demanding the global $U(1)$ symmetry imposes charge conservation at each vertex as in KK theory
\be\label{conservation}
\sum_{I=1}^n m_{i_I}=0
\ee
for every non-zero couplings where $I=1,..., n$ labels the legs of the vertex. Note that $m_i$ is really labelling the charge of $A^{i \mu}$ so we allow negative values $m_{-i} = - m_i$ with the understanding that the mass is $|m_i|$. The condition \eqref{conservation} ensures that the spectral conditions are satisfied \cite{Johnson:2020pny}. Since we have considered the EFT expansion of the action only up to  $1/\Lambda^4$ order, it is only consistent to calculate scattering amplitudes up to this order, which is precisely what we will do in Section \ref{sec:4pt} and Section \ref{sec:5pt} for the 4pt and 5pt functions respectively.

\section{4-point amplitudes}\label{sec:4pt} 

In this section we constrain the couplings of the Lagrangian by calculating 2-2 scattering amplitudes up to $1/\Lambda^4$ in the EFT expansion (for self consistency since we only add irrelevant operators up to $1/\Lambda^4$ in our action) and requiring the numerators to satisfy the 4pt colour-kinematic duality, $n_s+n_t+n_u=0$. We consider the scattering processes where the couplings are not fixed by 4d gauge invariance, for example, we do not consider $A^0A^0\rightarrow A^0A^0$.

\begin{figure}[t]
    \centering
    \includegraphics[width=15cm]{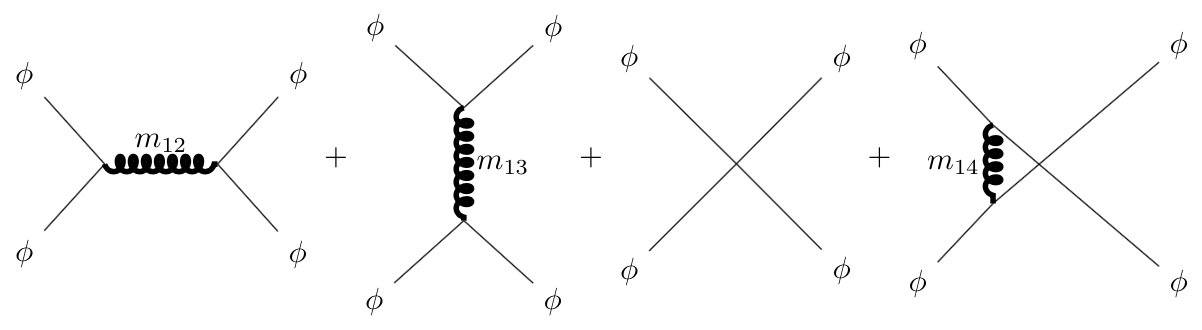}
    \caption{Feynman diagrams showing the $\phi\phi\rightarrow \phi\phi$ process. Note  that the bold curly line represents a gluon and $m_{ij}=0$.}
    \label{phi}
\end{figure}

\subsection{$\phi\phi\rightarrow\phi\phi$}
As a warm up example, let us first consider all the external states to be the massless scalars. By charge conservation, any exchanged state must be massless, and the only possibility for our chosen theory is the massless gluon. The Feynman diagrams for this process are in Fig.~\ref{phi} and the four point amplitude is found to be:
\ba
&&i{\cal A}_{\phi\phi\phi\phi}=\frac{i}{2 \Lambda ^4 s t (s+t)} \bigg(c_u s t (s-t) \big(-18
c_{\phi4} (s+t)^2+3 G_{0ss} (s+t) \left(\sqrt{2} g \Lambda ^2+6 G_{0ss} (s+t)\right) \nn \\
&&+g  \Lambda ^2 \left(g \Lambda ^2+3 \sqrt{2} G_{0ss} (s+t)\right)\big)-(s+t) \bigg(c_s t (s+2 t) \left(-18 c_{\phi4} s^2+g^2 \Lambda ^4-6 \sqrt{2} g G_{0ss} \Lambda ^2 s+18 G_{0ss}^2 s^2\right) \nn \\
&&-c_t s (2 s+t) \left(-18 c_{\phi4} t^2+g^2 \Lambda ^4-6 \sqrt{2} g G_{0ss} \Lambda ^2 t+18 G_{0ss}^2 t^2\right)\bigg)\bigg)
\ea
The BCJ relation gives the following condition
\begin{align*}
-\frac{i \left( c_{\phi4}- G_{0ss}^2\right) \left(3 s^2 t+2 s^3-3 s t^2-2 t^3\right)}{\Lambda ^4}=0,
\end{align*}
which is satisfied if 
\be
c_{\phi4}= G_{0ss}^2.
\ee
This simple example already illustrates the power of demanding CK duality. From a low energy EFT point of view, there is no reason for the Wilson coefficient for the quartic $(D \phi)^4$ operator to be associated with that for the cubic $D \phi D \phi F^0$ operator. Demanding CK duality enforces the relation `quartic coupling= cubic coupling squared'. Note that in this particular amplitude, no massive states are involved. We shall find that this general idea translates into constraints on nearly all the quartic operators from demanding CK duality. In the remainder, we will not include the explicit expression for the scattering amplitudes and BCJ relations as they are overly complicated, but shall give only the implied conclusion for the constraints on the coupling constants.

\subsection{General condition}

\begin{figure}[t]
    \centering
    \includegraphics[width=15cm]{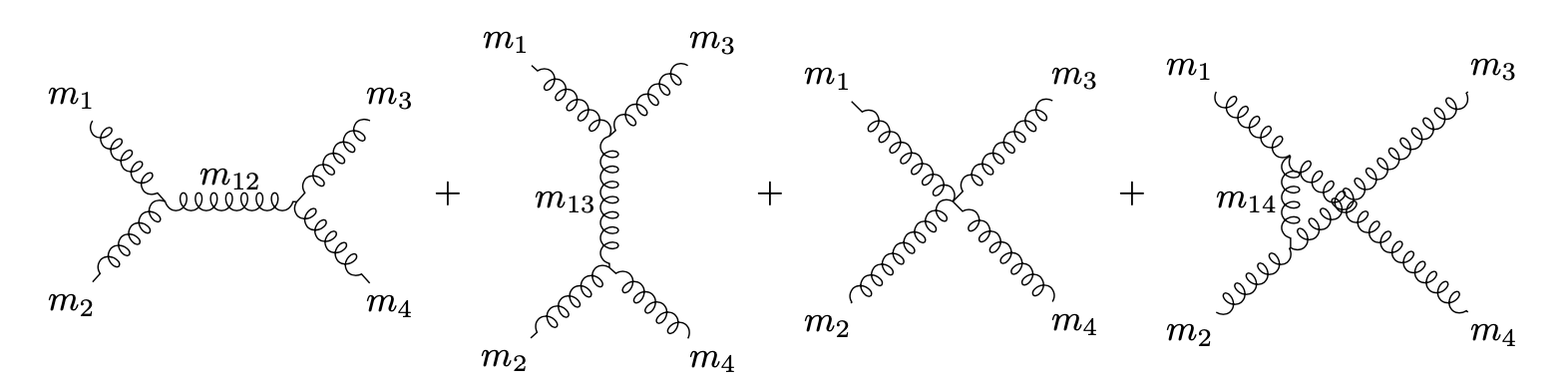}
    \caption{Feynman diagrams showing the $AA\rightarrow AA$ process for the general case. The curly lines represents massive spin-1 fields.}
    \label{fig:gn}
\end{figure}

We first calculate the 2-2 scattering amplitude of spin-1 fields with masses $m_I$, with $I=1,2,3,4$ labelling the amplitude legs, such that $\sum_{I=1}^4 m_{i_I}=0$, but for which charge conservation forbids the exchange of massless states. To avoid cumbersome notation we shall replace the charge label $i$ for the massive state with the leg label $I$ so that $m_{i_I}$ is denoted $m_I$. Similarly a coupling $G_{i_I i_J}$ is replaced with $G_{IJ}$. Furthermore, for cubic interactions between 3 charged states, for which the charge of the 3rd state is fixed by charge conservation, we shall drop the 3rd label. So for example $G_{i_1 i_2 i_3}$ for which $m_3=- m_1-m_2$ may be written in short hand as $G_{12}$. The interacting terms $\L_{AAA}$, $\L_{AAAA}$ and $\L_{AA\phi}$ lead to the following amplitude (see Fig.~\ref{fig:gn}):

\begin{equation}\label{general}
\begin{split}
    i {\cal A}_4\propto&\left(V^{AAA}_{g_{12}}+V^{AAA1}_{G_{12}}+V^{AAA2}_{\hat{G}_{12}}\right)\frac{i}{s-m^2_{12}}\left(V^{AAA}_{g_{34}}+V^{AAA1}_{G_{34}}+V^{AAA2}_{\hat{G}_{34}}\right)\\
   +&\left(V^{AAA}_{g_{13}}+V^{AAA1}_{G_{13}}+V^{AAA2}_{\hat{G}_{13}}\right)\frac{i}{t-m^2_{13}}\left(V^{AAA}_{g_{24}}+V^{AAA1}_{G_{24}}+V^{AAA2}_{\hat{G}_{24}}\right)\\
   +&\left(V^{AAA}_{g_{14}}+V^{AAA1}_{G_{14}}+V^{AAA2}_{\hat{G}_{14}}\right)\frac{i}{u-m^2_{14}}\left(V^{AAA}_{g_{23}}+V^{AAA1}_{G_{23}}+V^{AAA2}_{\hat{G}_{23}}\right)\\
   +&\left(V^{AAAA}_{g_{1234}}+V^{AAAA1}_{G_{1234}}+V^{AAAA2}_{\hat{G}_{1234}}+V^{AAAA1}_{c_{1234}}+V^{AAAA2}_{C_{1234}}\right),
\end{split}
\end{equation}
where $V^{ABC(D)}_{g_{ijk(l)}}$ represents the three or four point vertex with the relevant coupling, $g_{ijk(l)}$ or $G_{ijk(l)}$. The coupling $g_{ij}$ is a simplified notation for $g_{ijk}$ with $m_I+m_J+m_K=0$. Here we assume that the couplings are symmetric in all of their indices, for example $g_{jik}=g_{jik}=g_{kji}$ so that the $A_i A_j A_k$ vertex without $1/\Lambda^{2n}$ corrections has the same structure as the pure Yang-Mills three point vertex. Later we will explore what happens if we do not assume that. By finding the numerators of \eqref{general} and imposing the colour-kinematic duality, $n_s+n_t+n_u=0$, we find the following constraints on the couplings:
\begin{equation}\label{eq: BCJgeneral}
    \begin{split}
        &g_{1234}-\frac{18 m_1 m_2 m_3 m_4}{\Lambda^4}c_{1234}=g_{12}g_{34}=g_{13}g_{24}=g_{14}g_{23}=\frac{G^{2}_{1234}}{c_{1234}},\\
        &G_{1234}=G_{12}g_{34}=G_{13}g_{24}=G_{14}g_{23}=g_{12}G_{34}=g_{13}G_{24}=g_{14}G_{23},\\
        &G_{ij}=\hat{G}_{ij}, \quad G_{1234}=\hat{G}_{1234}, \quad c_{1234}=C_{1234},\\
        &c_{1234}=G_{12}G_{34}=G_{13}G_{24}=G_{14}G_{23} \, .
        \end{split}
\end{equation}

As mentioned in section \ref{sec:action}, $g_{1234}$ can be scale dependent as it is the combination of two terms coming from the compactification of $\L^{F^2}$ and $\L^{F^4}$. Therefore the coefficient of $1/\Lambda^4$ in the equation above, $c_{1234}$, does not have to be zero.

\subsection{$m_I+m_J=0$}

Further focussing on the case of 2-2 scattering amplitude of spin-1 fields with masses $m_I$, $I=1,2,3,4$, such that $m_I+m_J=0$ means that the exchange diagrams can now be mediated by the massless states, $A^0$ and $\phi$. First we consider the case where two pairs of masses add up to zero then the second case where four pairs of  masses add up to zero.

\begin{figure}[t]
    \centering
    \includegraphics[width=13cm]{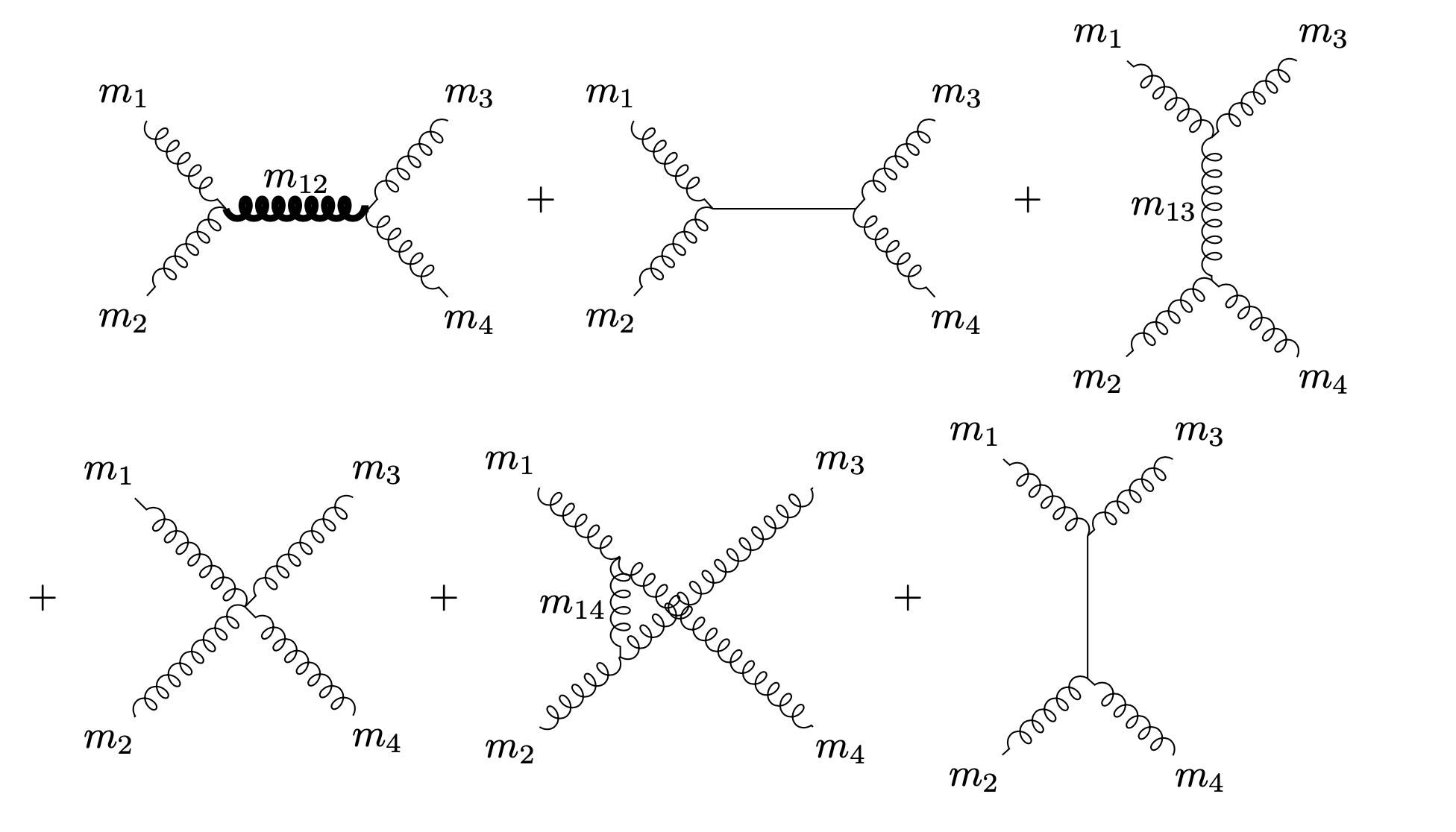}
    \caption{Feynman diagrams showing the $AA\rightarrow AA$ process for the first case of $m_I+m_J=0$, i.e.  $m_1+m_2=0$ and $m_3+m_4=0$. The bold curly line represents a gluon, where $m_{12}=0$ and the straight line a scalar field. The last diagram, where we have a scalar exchange in the $t$-channel contributes to the second case of $m_I+m_J=0$ where  $m_1=-m_2=m_3=-m_4$.}
    \label{fig:1}
\end{figure}

\paragraph{First case:} Without loss of generality, we consider $m_1+m_2=0$, which implies $m_3+m_4=0$. Moreover, we assume that none of the masses are individually zero. These conditions lead to an exchange of gluon and scalar in the $s$-channel and exchange of massive spin-1 field in the $t$-channel and the $u$-channel. Hence, from the interacting terms $\L_{AAA}$, $\L_{AAAA}$ and $\L_{AA\phi}$ we get the following amplitude (see Fig.~\ref{fig:1}):
:

\begin{equation}\label{case2}
\begin{split}
    i {\cal A}_4\propto&\left(V^{AAA_{0}}_{g}+V^{AAA_{0}}_{g_i}+V^{AAA_{0}}_{G_i}+V^{AAA_{0}}_{\hat{G}_i}\right)\frac{i}{s}\left(V^{AAA_{0}}_{g}+V^{AAA_{0}}_{g_i}+V^{AAA_{0}}_{G_i}+V^{AAA_{0}}_{\hat{G}_i}\right)\\
    +&\left(V^{AA\phi}_{g'_{12s}}+V^{AA\phi}_{G'_{12s}}\right)\frac{i}{s}\left(V^{AA\phi}_{g'_{34s}}+V^{AA\phi}_{G'_{34s}}\right)\\
   +&\left(V^{AAA}_{g_{13}}+V^{AAA1}_{G_{13}}+V^{AAA2}_{\hat{G}_{13}}\right)\frac{i}{t-m^2_{13}}\left(V^{AAA}_{g_{-1-3}}+V^{AAA1}_{G_{-1-3}}+V^{AAA2}_{\hat{G}_{-1-3}}\right)\\
   +&\left(V^{AAA}_{g_{1-3}}+V^{AAA1}_{G_{1-3}}+V^{AAA2}_{\hat{G}_{1-3}}\right)\frac{i}{u-m^2_{14}}\left(V^{AAA}_{g_{-13}}+V^{AAA1}_{G_{-13}}+V^{AAA2}_{\hat{G}_{-13}}\right)\\
   +&\left(V^{AAAA}_{g_{1-13-3}}+V^{AAAA1}_{G_{1-13-3}}+V^{AAAA2}_{\hat{G}_{1-13-3}}+V^{AAAA1}_{c_{1-13-3}}+V^{AAAA2}_{C_{1-13-3}}\right).
\end{split}
\end{equation}
By finding the numerators of \eqref{case2} and imposing the colour-kinematic duality, $n_s+n_t+n_u=0$, we find the following constraints on the couplings:
\begin{equation}\label{BCJcase2a}
    \begin{split}
       &g_{1-13-3}-\frac{18 m_1^2m_3^2}{\Lambda^4}c_{1-13-3}=g^2=g_{13}g_{-1-3}=g_{1-3}g_{-13},\\
       &G_{ij}=\hat{G}_{ij}, \quad g_i=g-G_i \frac{3\sqrt{2}m_i^2}{\Lambda^2}, \quad g'_{1-1s}g'_{3-3s}=g^2m_1m_3,\\
        &\hat{G}_{1-13-3}=G_{1-13-3}=G_{3}g=G_{1}g=G_{13}g_{-1-3}=g_{13}G_{-1-3}=G_{1-3}g_{-13}=g_{1-3}G_{-13}=\frac{g'_{1-1s}G'_{3-3s}}{m_1 m_3},\\
        &c_{1-13-3}=C_{1-13-3}=G_{1}G_{3}=G_{1-3}G_{-13}=G_{13}G_{-1-3}.
    \end{split}
\end{equation}
When $m_4=m_3=0$ we find that 
\begin{equation}\label{BCJcase2a m3m4 zero}
    \begin{split}
        &G_{i}=G,\quad c_{1-100}=C_{1-100}=G^2 \, .
    \end{split}
\end{equation}
\begin{figure}[t]
    \centering
    \includegraphics[width=15cm]{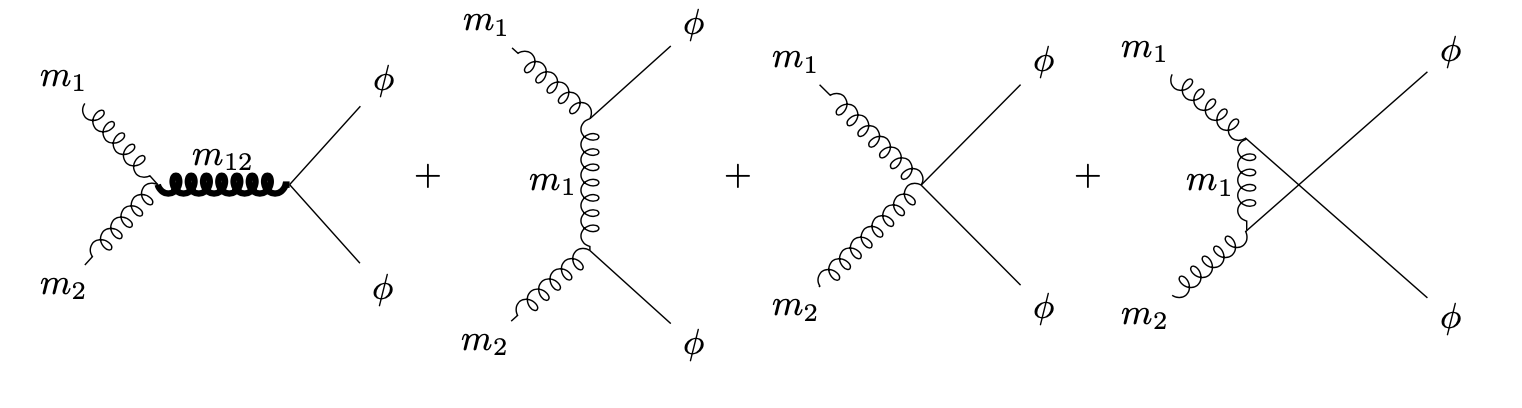}
    \caption{Feynman diagrams showing the $AA\rightarrow \phi\phi$ process for $m_I+m_J=0$ case. The bold curly line represents a gluon, where $m_{12}=0$ and the straight line a scalar field.}
    \label{fig:3}
\end{figure}
\paragraph{Second case:}
Now we consider four pairs of masses to be zero, for example, $m_1+m_2=m_1+m_3=m_3+m_4=m_2+m_4=0$ and $m_1\neq 0$. These conditions lead to an exchange of gluon and scalar in the $s$ and $t$ channels and exchange of massive spin-1 field in the $u$-channel. Hence, from the interacting terms $\L_{AAA}$, $\L_{AAAA}$ and $\L_{AA\phi}$ we get the following amplitude (see Fig.~\ref{fig:1}):
:
\begin{equation}\label{case21}
\begin{split}
    i {\cal A}_4\propto&\left(V^{AAA_{0}}_{g}+V^{AAA_{0}}_{g_i}+V^{AAA_{0}}_{G_i}\right)\frac{i}{s}\left(V^{AAA_{0}}_{g}+V^{AAA_{0}}_{g_i}+V^{AAA_{0}}_{G_i}\right)\\
    +&\left(V^{AA\phi}_{g'_{1-1s}}+V^{AA\phi}_{G'_{1-1s}}\right)\frac{i}{s}\left(V^{AA\phi}_{g'_{-11s}}+V^{AA\phi}_{G'_{-11s}}\right)\\
   +&\left(V^{AAA}_{g_{1-1}}+V^{AAA1}_{G_{1-1}}+V^{AAA2}_{\hat{G}_{1-1}}\right)\frac{i}{t}\left(V^{AAA}_{g_{-11}}+V^{AAA1}_{G_{-11}}+V^{AAA2}_{\hat{G}_{-11}}\right)\\
    +&\left(V^{AA\phi}_{g'_{1-1s}}+V^{AA\phi}_{G'_{1-1s}}\right)\frac{i}{t}\left(V^{AA\phi}_{g'_{-11s}}+V^{AA\phi}_{G'_{-11s}}\right)\\
   +&\left(V^{AAA}_{g_{11}}+V^{AAA1}_{G_{11}}+V^{AAA2}_{\hat{G}_{11}}\right)\frac{i}{u-m^2_{11}}\left(V^{AAA}_{g_{-1-1}}+V^{AAA1}_{G_{-1-1}}+V^{AAA2}_{\hat{G}_{-1-1}}\right)\\
   +&\left(V^{AAAA}_{g_{1-1-11}}+V^{AAAA1}_{G_{1-1-11}}+V^{AAAA2}_{\hat{G}_{1-1-11}}+V^{AAAA1}_{c_{1-1-11}}+V^{AAAA2}_{C_{1-1-11}}\right).
\end{split}
\end{equation}
By finding the numerators of \eqref{case21} and imposing the colour-kinematic duality, $n_s+n_t+n_u=0$, we find the following constraints on the couplings \footnote{We found another set of solution where the quartic operators $G_{1-1-11}=\hat{G}_{1-1-11}=0$ but $G_{14}\neq \hat{G}_{14}$ which is incompatible with the result obtained from the general case.}:
\begin{equation}\label{BCJcase2a1}
    \begin{split}
       &g_{1-1 -1 1}-\frac{18 m_1^4}{\Lambda^4}c_{1-1 -1 1}=g^2=g_{1 1}g_{-1 -1}, \quad g_i=g-G_i \frac{3\sqrt{2}m_i^2}{\Lambda^2}, \quad g'_{1-1s}g'_{-1 1s}=g^2 m_1^2,\\
        &\hat{G}_{1-1 -11}=G_{1-1 -11}=G_{-1}g=G_{1}g=G_{1 1}g_{-1 -1}=g_{1 1}G_{-1 -1}=\frac{g'_{1-1s}G'_{-1 1s}}{m_1^2},\\
        &c_{1-1 -1 1}=C_{1-1 -1 1}=G_{1}G_{-1}=G_{1-1}G_{-1-1}.
    \end{split}
\end{equation}

\subsubsection{$AA\rightarrow \phi\phi$}
Now we consider the case where $m_3=m_4=0$ and we calculate the $AA\rightarrow \phi\phi$ amplitude. A gluon is exchanged in the $s$-channel and a massive spin-1 field in the $t$ and $u$-channel. From the terms $\L_{AAA}$, $\L_{AA\phi\phi}$, $\L_{AA\phi}$ and $\L_{\phi\phi A}$ we have (see Fig.~\ref{fig:3}):
\begin{equation}\label{m3m4}
\begin{split}
    i{\cal A}_4\propto&\left(V^{AAA_{0}}_{g}+V^{AAA_{0}}_{g_i}+V^{AAA_{0}}_{G_i}\right)\frac{i}{s}\left(V^{A_{0}\phi\phi}_{g}+V^{A_{0}\phi\phi}_{G_{0ss}}\right)\\
   +&\left(V^{AA\phi}_{g'_{1-1s}}+V^{AA\phi}_{G'_{1-1s}}\right)\frac{i}{t-m^2_{1}}\left(V^{AA\phi}_{g'_{1-1s}}+V^{AA\phi}_{G'_{1-1s}}\right)\\
   +&\left(V^{AA\phi}_{g'_{1-1s}}+V^{AA\phi}_{G'_{1-1s}}\right)\frac{i}{u-m^2_{1}}\left(V^{AA\phi}_{g'_{1-1s}}+V^{AA\phi}_{G'_{1-1s}}\right)\\
   +&\left(V^{AA\phi\phi}_{g_{1-100}}+V^{AA\phi\phi1}_{G_{1-1ss}}+V^{AA\phi\phi2}_{\hat{G}_{1-1ss}}+V^{AA\phi\phi1}_{c_{1-1ss}}+V^{AA\phi\phi2}_{c^{(2)}_{1-1ss}}+V^{AA\phi\phi3}_{c^{(3)}_{1-1ss}}\right).
\end{split}
\end{equation}
Colour-kinematics duality puts the following constraints on the couplings:
\begin{equation}\label{BCJcase2b}
    \begin{split}
        &g_{i-iss}=g^2, \quad G_{i}= \frac{2{g'}^2_{i-is}-m_1^2 g(g+g_i)}{3\sqrt{2}g m_{1}^{2}}\frac{\Lambda^2}{m^{2}_{1}}, \quad c_{i-iss}=G_{0ss}G_{i}, \quad c^{(2)}_{i-iss}=\frac{G^{'2}_{i-is}}{m_1^2}\\
        &c^{(3)}_{i-iss}=G_{0ss}G_{i}+\frac{\sqrt{2}\Lambda^2}{6m_1^2}g\left(G_{0ss}-G_{i}\right) , \quad \hat{G}_{i-iss}=gG_{i} \, \quad G_{i-iss}=-gG_{i}+\frac{2g^{'2}_{i-is} G_{0ss}}{g m_1^2}-\frac{3\sqrt{2}G^{'2}_{i-is}}{\Lambda^2}.
    \end{split} 
\end{equation}

\begin{figure}[t]
    \centering
    \includegraphics[width=15cm]{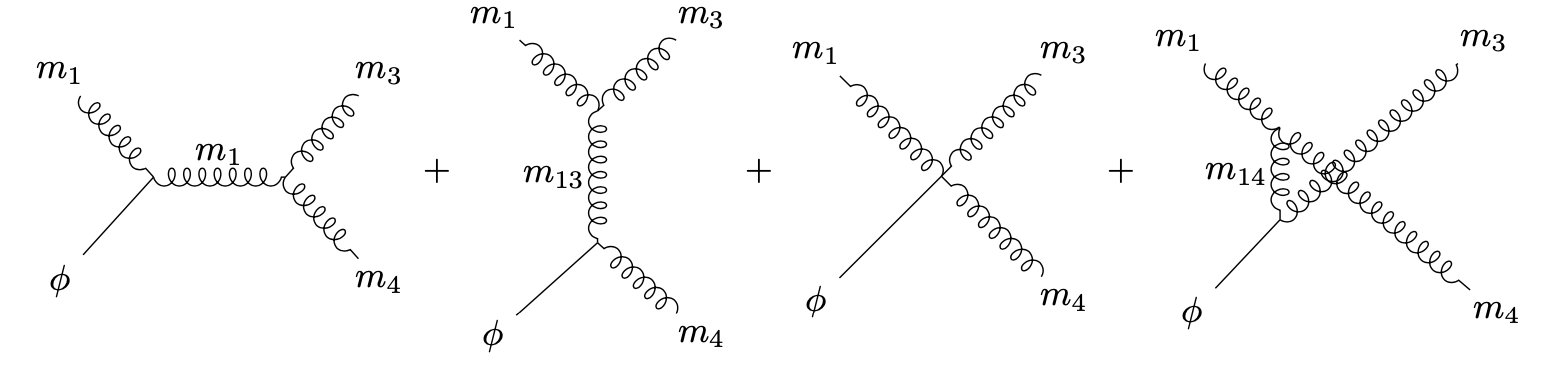}
    \caption{Feynman diagrams showing the $A\phi\rightarrow AA$ process for $m_I=0$ case.}
    \label{fig:4}
\end{figure}

\subsection{$m_I=0$}

In this section, we  calculate 2-2 scattering amplitude of one massless external state and three massive spin-1 fields satisfying $m_I+m_J+m_K=0$. We consider the massless state to be a scalar with $m_2=0$ (note that the case where the massless state is a gluon reproduces the same results as in the general case and we will give the constraints at the end of this section for completeness). The amplitude for this process is the following (see Fig.~\ref{fig:4}):

\begin{equation}\label{scalar}
\begin{split}
    i{\cal A}_4\propto&\left(V^{AA\phi}_{g'_{1-1s}}+V^{AA\phi}_{G'_{1-1s}}\right)\frac{i}{s-m^2_{1}}\left(V^{AAA}_{g_{34}}+V^{AAA1}_{G_{34}}+V^{AAA2}_{\hat{G}_{34}}\right)\\
   +&\left(V^{AAA}_{g_{13}}+V^{AAA1}_{G_{13}}+V^{AAA2}_{\hat{G}_{13}}\right)\frac{i}{t-m^2_{13}}\left(V^{AA\phi}_{g'_{4-4}s}+V^{AA\phi}_{G'_{4-4}s}\right)\\
   +&\left(V^{AAA}_{g_{14}}+V^{AAA1}_{G_{14}}+V^{AAA2}_{\hat{G}_{14}}\right)\frac{i}{u-m^2_{14}}\left(V^{AA\phi}_{g'_{3-3s}}+V^{AA\phi}_{G'_{3-3s}}\right)\\
   +&\left(V^{AAA\phi1}_{\hat{G}_{134s}}+V^{AAA\phi2}_{G_{134s}}+V^{AAA\phi1}_{c_{134s}}+V^{AAA\phi2}_{C_{134s}}\right),
\end{split}
\end{equation}
Note that in this case we have $m_1+m_3+m_4=0$, so:
\begin{equation}
\begin{split}
    g_{34(-3-4)}&=g_{341}, \quad g_{13(-1-3)}=g_{134}, \quad g_{14(-1-4)}=g_{143},\\
    G_{34(-3-4)}&=G_{341}, \quad   G_{13(-1-3)}=G_{134}, \quad G_{14(-1-4)}=G_{143},
\end{split}
\end{equation}
where we assume that $g_{ijk}$ and $G_{ijk}$ are fully symmetric in their indices and $(\pm I\pm J)$ represents $\pm m_{I}\pm m_{J}$. Hence, in our simplified notation we have  $g_{34}=g_{13}=g_{14}$ and $G_{34}=G_{13}=G_{14}$ in this case. Imposing the colour-kinematic duality on $A\phi\rightarrow AA$, we get the following constraints:
\begin{equation}\label{eq: bcjcase3b HO}
    \begin{split}
        &\frac{g'_{4-4s}}{g'_{3-3s}}=\frac{m_4}{m_3}, \quad \frac{g'_{1-1s}}{g'_{3-3s}}=\frac{m_1}{m_3},  \quad G_{34}=\hat{G}_{34}\\
        &\frac{G_{13}g_{4-4}'}{m_4}=\frac{G_{14}g_{3-3}'}{m_3}=\frac{g_{13}G_{4-4s}'}{m_4}=\frac{g_{14}G_{3-3s}'}{m_3}=\hat{G}_{134s}=G_{134s},\\
        &c_{134s}=C_{134s}=\frac{G_{1-1s}G_{34}}{m_1}=\frac{G_{4-4s}G_{13}}{m_4}=\frac{G_{3-3s}G_{14}}{m_3}.
    \end{split}
\end{equation}
The conditions on couplings where one external state ($m_2=0$) is a gluon can be derived from \eqref{eq: BCJgeneral} by setting $g_{12}=g_{24}=g_{23}=g$ and $g_{1234}=g^2$ which leads to $g_i=g-G_i \frac{3\sqrt{2}m_i^2}{\Lambda^2}$ and $g_{34} g=g^2$ so $g_{ij}=g$ for arbitrary $i,j$, and $G_{1}g=G_{13}g$ which, when combined with \eqref{eq: BCJgeneral} and \eqref{BCJcase2a m3m4 zero}, implies $G_i=G_{ij}=\hat{G}_{ij}=G$, i.e. we fix all cubic couplings except for $G_{0ss}$. The two quartic $AAAA^0$ couplings coming from $\L^{F^3}_{AAA1}$ and $\L^{F^3}_{AAA2}$ in \eqref{eq:HO} are both equal to $G_{34}g$ by gauge invariance so BCJ relation does not impose an additional constraint on them.\\

\textbf{Summary of results}: Combining all of the constraints obtained from different cases and different processes we obtain the results summarized in table~\ref{KK result}. These match the values of couplings of the 4d KK theory of 5d Yang-Mills with coupling $\sqrt{2\pi R} g$ plus $\frac{\sqrt{2\pi R}G}{\Lambda^2}tr(F^3)-\frac{9\pi R G^2}{8\Lambda^4}tr([F_{\mu\nu},F_{\alpha\beta}][F^{\mu\nu},F^{\alpha\beta}])$ operators, where $R$ is the radius of the $S^1$, if $G_{0ss}=G$, but this is not fixed from 4pt processes. The reason for this is that $\L^{F^3}_{A\phi\phi}$ vertex in \eqref{eq:HO} is zero when $A$ is on-shell. This means that the only 4pt processes that can constrain $G_{0ss}$ are $\phi\phi\rightarrow\phi\phi$ and $AA\rightarrow\phi\phi$ but as we saw in the previous sections there is a freedom between $G_{0ss}$ and quartic $\Lambda^4$ coefficients in both of these processes. Therefore, this coupling can only be fixed by 5pt scattering.

\begin{table}[t]
    \centering
    \begin{tabular}{|c|c|c|c|c|c|}
    \hline
         $ $ & coefficient & CK constrained value  &  $ $ & coefficient & CK constrained value \\
         \hline\hline
         $\L_{AAA}$ & $g_{ijk}$ & $ g$ & $\L^{F^3}_{AA\phi\phi 1}$ & $G_{ijss}$& $g(2G_{0ss}-G)-\frac{3\sqrt{2}m_i^2G^2 }{\Lambda^2}$ \\
         \hline
         $\L_{AA\phi}$& $g'_{ijs}$ & $m_{i}g$ &      $\L^{F^3}_{AA\phi\phi 2}$ & $\hat{G}_{ijss}$ & $gG$ \\
         \hline
         $\L_{AAAA}$ & $g_{ijkl}$ & $g^2+\frac{18 m_i m_j m_k m_l}{\Lambda^4}G^2$ & $\L^{F^3}_{AAA\phi 1}$ & $\hat{G}_{ijks}$ & $gG$   \\
         \hline
         $\L_{AA\phi\phi}$ & $g_{ijss}$ & $ g^2$ & $\L^{F^3}_{AAA\phi 2}$ & $G_{ijks}$ & $gG$ \\
         \hline
         $ \L^{F^4}_{AAF^0}$ & $g_i$ & $g_i=g-G \frac{3\sqrt{2}m_i^2}{\Lambda^2}$ & $\L^{F^4}_{AAAA1}$ & $c_{ijkl}$ & $G^2$ \\
\hline
$ \L^{F^3}_{AAF^0}$ & $G_i$ & $G_{i}=G$ & $\L^{F^4}_{AAAA2}$ & $C_{ijkl}$  & $G^2$ \\
\hline 
$\L^{F^3}_{AAA1}$ & $G_{ijk}$ & $G$ & $\L^{F^4}_{AAA\phi1}$ & $c_{ijks}$ & $G^2$\\
         \hline
         $\L^{F^3}_{AAA2}$ & $\hat{G}_{ijk}$& $G$ & $\L^{F^4}_{AAA\phi2}$ & $C_{ijks}$ & $G^2$ \\
         \hline
         $\L^{F^3}_{AA\phi}$ & $G'_{ijs}$& $m_i G$  & $\L^{F^4}_{AA\phi\phi1}$ & $c_{ijss}$ & $G_{0ss}G$ \\
         \hline
         $\L^{F^3}_{A\phi\phi}$ & $G_{0ss}$& $\text{not constrained}$ & $ \L^{F^4}_{AA\phi\phi2}$ & $c^{(2)}_{ijss}$  & $G^2$ \\
         \hline
         $\L^{F^3}_{AAAA1}$ & $G_{ijkl}$ & $gG$ & $ \L^{F^4}_{AA\phi\phi3}$ & $c^{(3)}_{ijss}$  & $G_{0ss}G+\frac{\sqrt{2}\Lambda^2}{6m_i^2}g\left(G_{0ss}-G\right)$ \\
         \hline
         $\L^{F^3}_{AAAA2}$ & $\hat{G}_{ijkl}$& $gG$ & $ \L^{F^4}_{\phi\phi\phi\phi}$ & $c_{\phi 4}$ & $G^{2}_{0ss}$ \\
         \hline
       \end{tabular}
    \caption{Coefficients of the interactions constrained by the demands of colour-kinematics duality.}
    \label{KK result}
\end{table}

\section{5-point amplitudes}\label{sec:5pt}

In this section we consider different 5pt scattering amplitudes of external massive and massless fields and find the CK constraints on contact couplings imposed by the BCJ relations. Since we know from the previous section that all 3pt and 4pt couplings, except for $G_{0ss}$, are fixed to be that of KK theory, the only remaining freedom at 5pt is that from the as yet undetermined $G_{0ss}$, and the additional $1/\Lambda^2$ and $1/\Lambda^4$ suppressed quintic operators. We focus on the cases where we have multiple quintic contact terms that are not fixed by gauge invariance. For instance, the $A^0AAAA$ operators are clearly fixed by determining the $AAAA$ interactions by gauge invariance. As we will show below the remaining undetermined cubic coupling $G_{0ss}$ is fixed by considering CK duality for $\phi\phi AAA$ scattering.  In addition we shall find that the quintic contact terms are fixed in terms of a single coupling constant. This result is perhaps not so surprising since there are more BCJ relations to be satisfied, more independent contractions of polarizations and momenta, yet fewer overall free coefficients for quintic contact terms.

\subsection{General case}

\begin{figure}[H]
    \centering
    \includegraphics[width=10cm]{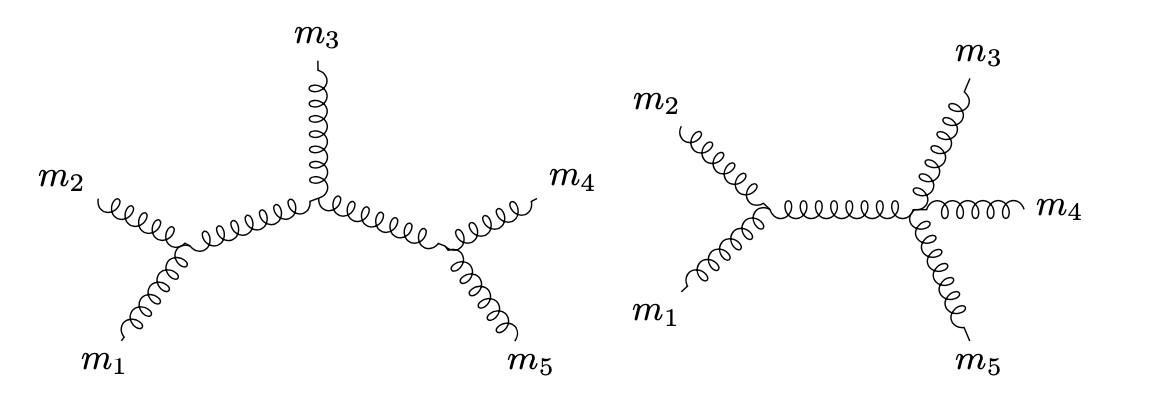}
    \caption{Five point diagrams for the general case.}
    \label{fig: 5pt general}
\end{figure}

We consider the 5-point scattering amplitude of five external spin-1 fields of mass $m_I$, $I=1,2,3,4,5$ such that $m_1+m_2+m_3+m_4+m_5=0$ and there are no $I,J$ such that $m_I+m_J=0$. Only massive spin-1 fields are exchanged in the 25 diagrams contributing to the amplitude (note that we have 15 kinematic factors as we can absorb the quartic contact terms into cubic diagrams). The two types of diagrams are shown in Fig.~\ref{fig: 5pt general} and we obtained all 25 diagrams by relabelling the external states of these two. We then calculate the kinematic numerators and 5pt BCJ relations given in \eqref{eq:bcj1} - \eqref{eq:bcj4} and require them to be zero which gives us the constraints on quintic contact term coefficients. We use \eqref{KK result} for the cubic and quartic couplings and find that all the three coefficients, $G_{ijklm}, c_{ijklm}$ and $C_{ijklm}$ are fixed by $(\epsilon_1\cdot\epsilon_2) (\epsilon_4\cdot\epsilon_5 )(\epsilon_3\cdot p_1)$ term in \eqref{eq:bcj1} to be of the same values as in KK theory: 
\be \label{eq: AAAAA constraints}
G_{ijklm}=g^2 G, \; c_{ijklm}=C_{ijklm}=gG^2.
\ee

\subsection{$m_1+m_2=0$}

\begin{figure}[H]
    \centering
    \includegraphics[width=14cm]{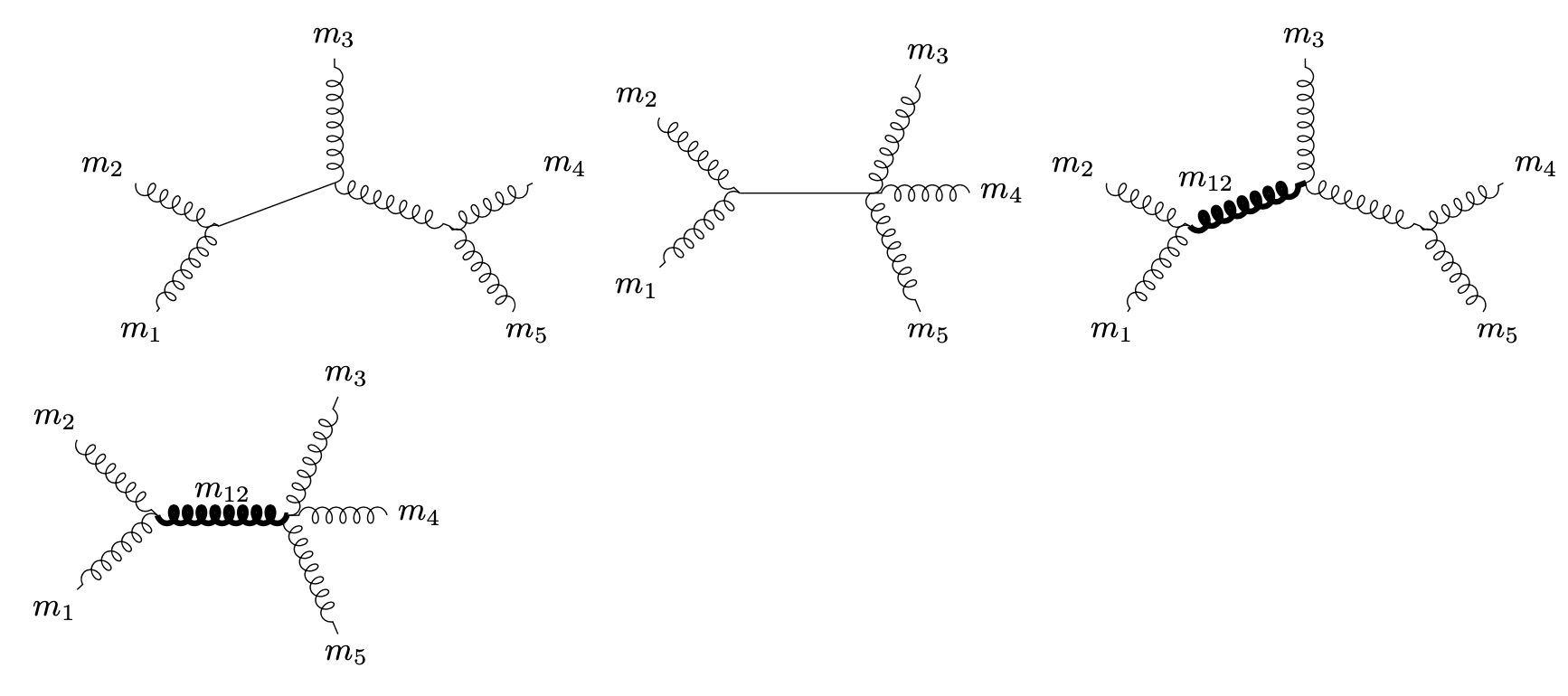}
    \caption{Five point diagrams for the general case when $m_1+m_2=0$. The bold curly line represents a massless gluon, where $m_{12}=0$ and the straight line a massless scalar.}
    \label{fig: AAAAA m12}
\end{figure}

Next we consider the 5-point scattering amplitude of five external spin-1 field such that $m_1+m_2=0$. In this case, 6 out of the 30 diagrams have a massless particle exchange, a gluon and a scalar. These diagrams are shown in Fig.~\ref{fig: AAAAA m12}.
In exactly the same way as in previous case, \eqref{eq:bcj1} forces the quintic couplings to be as in \eqref{eq: AAAAA constraints}.

\subsection{$AAAAA$ without $\Lambda^{-2n}$ operators}

\begin{figure}[t]
    \centering
    \includegraphics[width=10cm]{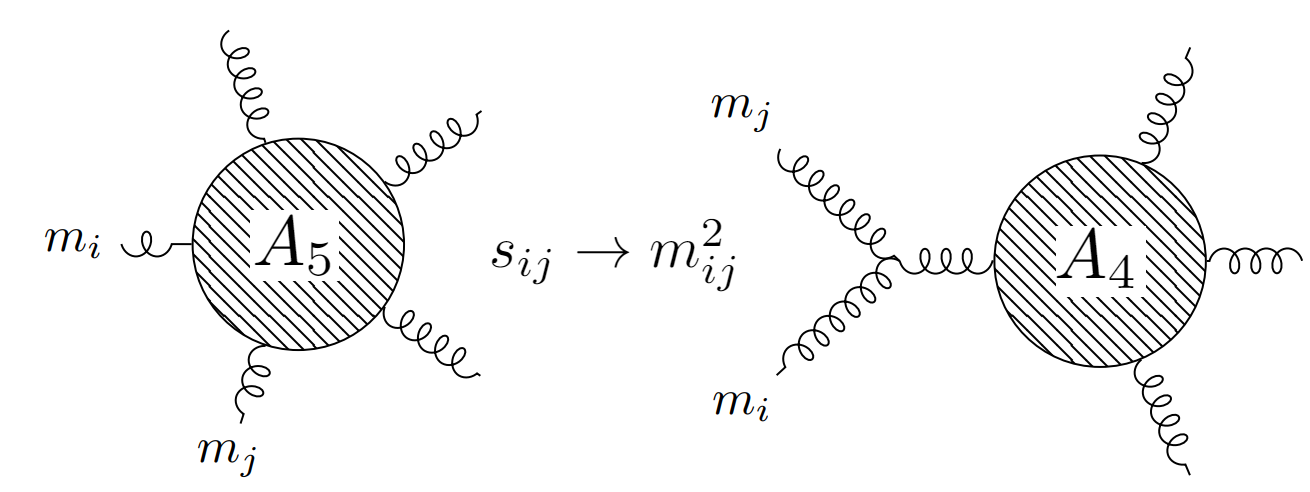}
    \caption{Factorization limits of the 5-point amplitude. By only imposing the 4-point BCJ relations on $A_4$ for all possible factorizations, the 5-point BCJ can be satisfied}
    \label{fig: facto}
\end{figure}

Let us temporarily take a step back, and consider the $AAAAA$ amplitude at leading order in the EFT expansion $\Lambda^0$. We find that imposing the constraints from $AA\rightarrow AA$ scattering is enough for the BCJ relation to hold at 5-point, at least to this order in the EFT expansion. This can be seen by taking the factorization limits of the 5-point amplitude and imposing  BCJ relations on the sub 4-point amplitudes (see Fig.~\ref{fig: facto}). To show our procedure of checking the BCJ at 5-point, consider the amplitude for this process as:
\begin{equation}
   {\cal A}_{5}=g_{24}g_{15}g_{3(2+4)}\frac{(...)}{D_{15}D_{24}}+ g_{24}g_{35}g_{1(2+4)}\frac{(...)}{D_{24}D_{35}}+ g_{24}g_{135(2+4)}\frac{(...)}{D_{24}}+\text{other contributions}
\end{equation}
where $(...)$ represents the contraction of polarizations, momenta and the colour factors. From \eqref{eq: BCJgeneral}, we express the quartic coupling $g_{135(24)}$ as a product of two cubic diagrams:
\begin{equation}
\begin{split}
    g_{135(24)}&=g_{35}g_{1(3+5)},\\
    &=g_{15}g_{3(1+5)},\\
    &=g_{13}g_{5(1+3)}.
\end{split}
\end{equation}

We also express the  9 remaining contact couplings in terms of products of cubic ones. We combine all 30 equalities (3 equalities per contact coupling), simultaneously solve them and we get 24 constraints. Once these constraints imposed on the kinematic factors, the 5-point BCJ relation is satisfied. We conclude that for our theory, the BCJ relations at 5-point are satisfied and can be obtained from the coupling conditions derived at 4-point. This is consistent with the fact that BCJ relation can be proved recursively using BCFW recursion for theories in which amplitudes can be constructed that way. Of course this will not be true if we add $1/\Lambda^{2n}$ corrections because the 5pt contact term couplings there are independent from 3 and 4pt couplings. 

\subsection{$\phi\,\phi AAA$}
\begin{figure}[t]
    \centering
    \includegraphics[width=12cm]{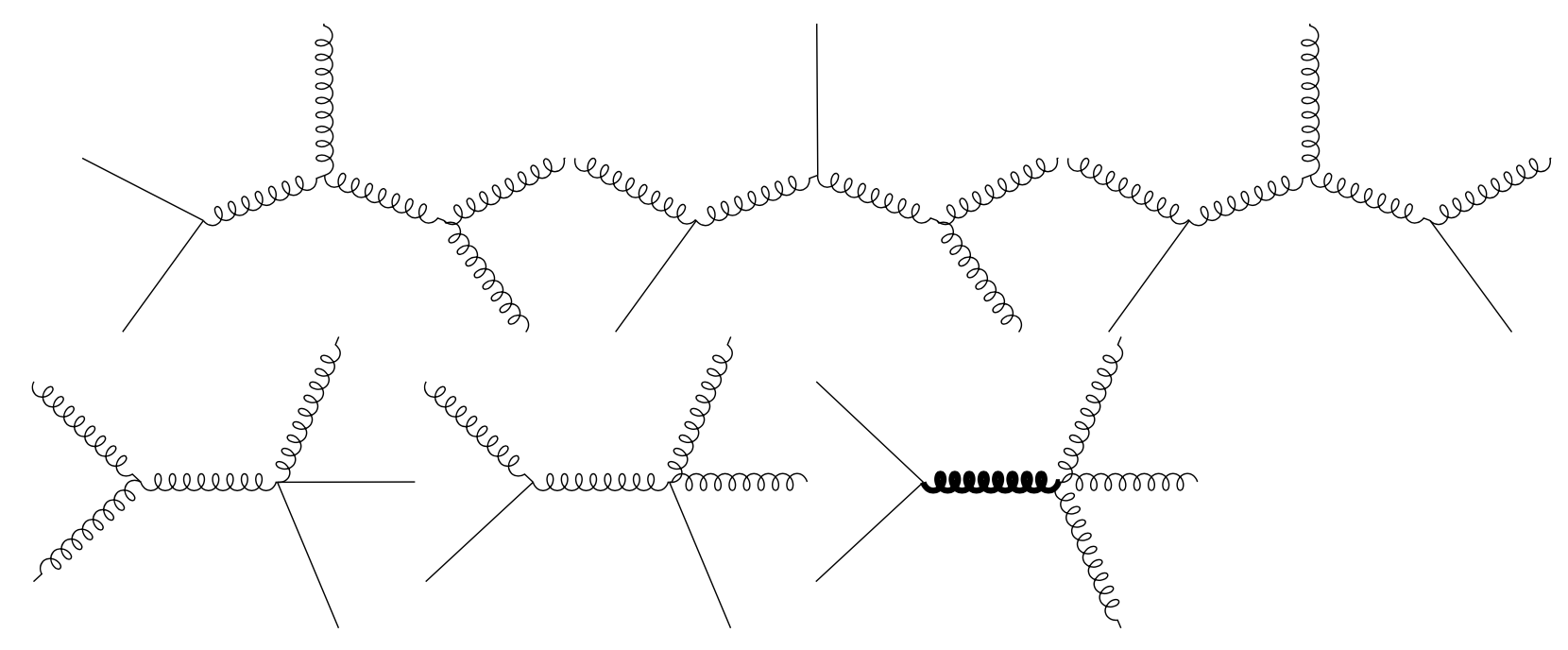}
    \caption{The six types of diagrams for $\phi\,\phi AAA$ process. The bold curly line represents a gluon and the straight line a scalar field.}
    \label{fig: ssAAA}
\end{figure}

Now we consider two massless scalars and three massive vectors with the diagrams shown in Fig.~\ref{fig: ssAAA}. By inspection of the $(\epsilon_3\cdot p_1) (\epsilon_5\cdot p_2) (\epsilon_4\cdot p_2$), $(\epsilon_3\cdot\epsilon_4 )(\epsilon_5\cdot p_2 )$ and $(\epsilon_5\cdot\epsilon_4 )(\epsilon_3\cdot p_2 )$ terms in \eqref{eq:bcj1} and \eqref{eq:bcj4} we are able to finally fix $G_{0ss}=G$ (which fixes all 4pt couplings as in \eqref{KK result}) and the four quintic couplings to be of their KK values:
\be
G_{ijkss}=\hat{G}_{ijkss}=g^2 G, \; c_{ijkss}=C_{ijkss}=\hat{C}^{(3)}_{ijkss}=\hat{C}^{(4)}_{ijkss}=\hat{C}^{(5)}_{ijkss}=gG^2.
\ee

\subsection{$\phi AAAA$}
\begin{figure}[t]
    \centering
    \includegraphics[width=12cm]{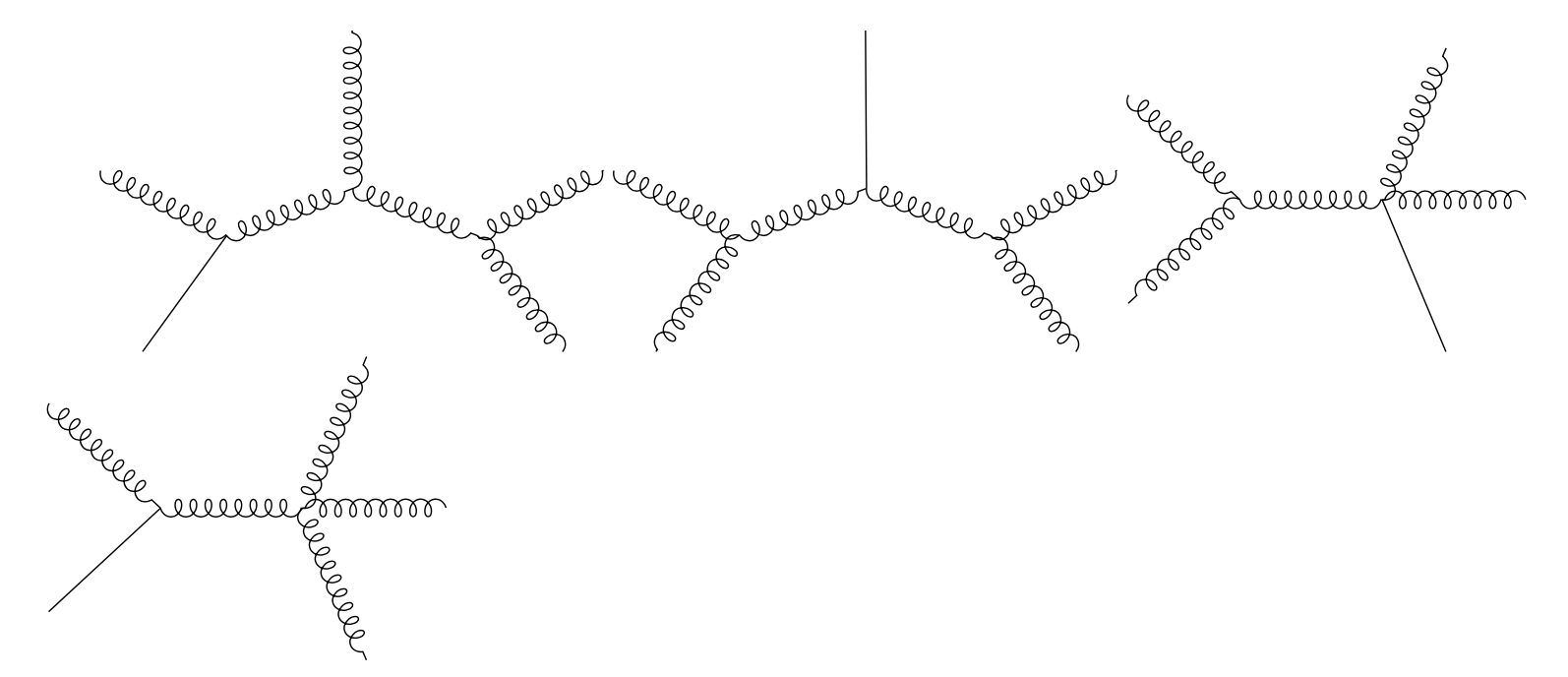}
    \caption{The four types of diagrams for $\phi AAAA$ process.The straight line is a scalar field.}
    \label{fig: sAAAA}
\end{figure}

Now we consider the first state to be a scalar, and the remaining states to be massive vectors. Let us first assume there are no pairs of vectors for which $m_I+m_J\neq0$. Then the four types of diagrams are shown in Fig.~\ref{fig: sAAAA}. Inspecting the $(\epsilon_2\cdot\epsilon_3)(\epsilon_4\cdot\epsilon_5)$ and $(\epsilon_2\cdot\epsilon_3 )(\epsilon_4\cdot p_1 )(\epsilon_5\cdot p_2)$ terms in \eqref{eq:bcj1} and \eqref{eq:bcj3} fix the four quintic couplings to be of their KK values:

\be \label{eq: sAAAA constraints}
G_{ijkls}=\hat{G}_{ijkls}=g^2 G, \; c_{ijkls}=C_{ijkls}=\hat{C}_{ijkls}=gG^2.
\ee
From this we can consider two special cases where either a single or a double pair of massive vectors satisfy $m_I+m_J\neq0$.

\subsubsection{ $m_I+m_J=0$}

\begin{figure}[t]
    \centering
    \includegraphics[width=12cm]{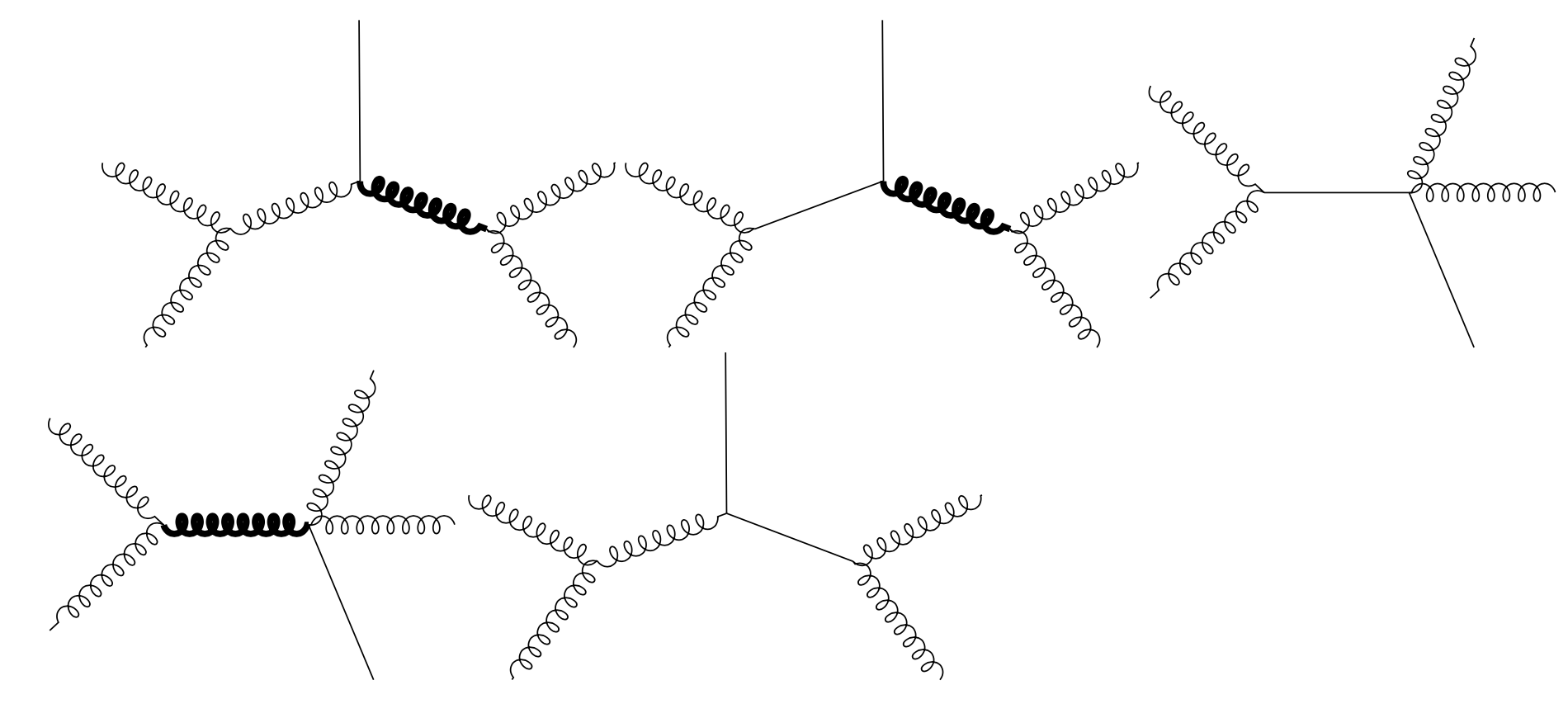} 
    \caption{The five types of diagrams for $\phi AAAA$ process when $m_i+m_j=0$. The bold curly line represents a gluon and the straight line a scalar field.}
    \label{fig: sAAAA m12}
\end{figure}

\paragraph{Case 1:} Consider again one scalar and four vectors but only one pair of masses add up to zero, for example, we consider  $m_2+m_3=m_4+m_5=0$. There are more diagrams now which are shown in Fig.~\ref{fig: sAAAA m12}. We use the constraint $G_{0ss}=G$ obtained from $\phi\,\phi AAA$. Now inspecting the $(\epsilon_2\cdot\epsilon_3)( \epsilon_4\cdot\epsilon_5)$ and $(\epsilon_2\cdot\epsilon_3)( \epsilon_4\cdot p_2)( \epsilon_5\cdot p_3)$ terms in \eqref{eq:bcj2} fixes the four quintic couplings to be of their KK values given in \eqref{eq: sAAAA constraints}.

\paragraph{Case 2:} We consider once again one scalar and four vectors but their masses now satisfy $m_2=-m_3=-m_4=m_5$. The diagrams still look like the ones in Fig.~\ref{fig: sAAAA m12} but now we have more channels with scalar or massless gluon exchanges. Just as before $(\epsilon_2\cdot\epsilon_3)( \epsilon_4\cdot\epsilon_5)$ and $(\epsilon_2\cdot\epsilon_3)( \epsilon_4\cdot p_2)( \epsilon_5\cdot p_3)$ terms in \eqref{eq:bcj2} fix the four quintic couplings to be of their KK values given in \eqref{eq: sAAAA constraints}.

\begin{figure}[t]
    \centering
    \includegraphics[width=12cm]{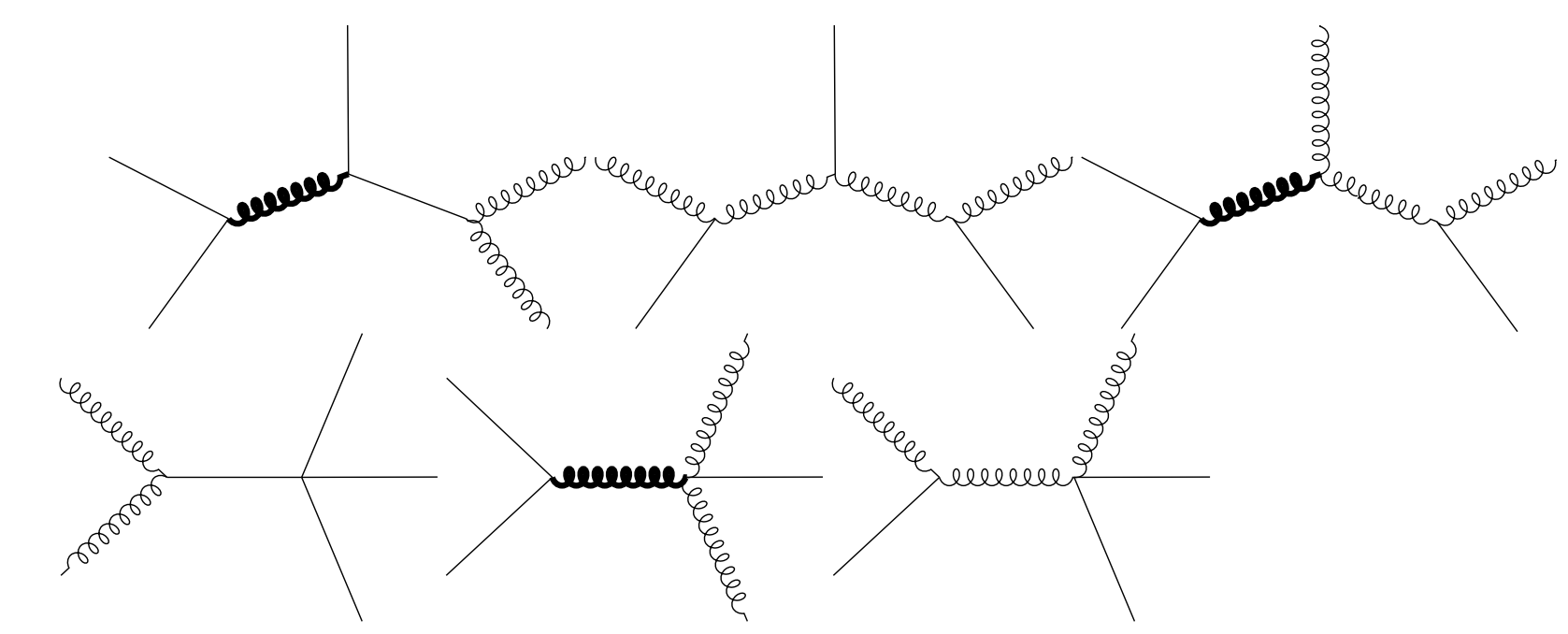}
    \caption{The six types of diagrams for $\phi\,\phi\,\phi AA$ process. The bold curly line represents a gluon and the straight line a scalar field.}
    \label{fig: sssAA}
\end{figure}

\subsection{$\phi\,\phi\,\phi AA$}
Finally we consider three scalars and two vectors with the diagrams shown in Fig.~\ref{fig: sssAA}. We now make use of the constraint $G_{0ss}=G$ obtained from $\phi\,\phi AAA$. In this case we only have the $1/\Lambda^4$ quintic operators and by inspecting $\epsilon_4\cdot\epsilon_5$ and $(\epsilon_5\cdot p_2 )(\epsilon_4\cdot p_1 )$ terms in \eqref{eq:bcj1} we fix their coefficients to be:
\be
c_{ijsss}=c^{(2)}_{ijsss}=c^{(3)}_{ijsss}=gG^2.
\ee

\section{Non-symmetric couplings}\label{nonsym}
In the previous sections we assumed the couplings $g_{ijk}$, $g_{ijkl}$ and $G_{ijkl}$ to be fully symmetric in all of the indices. This is of course what we obtain from KK reduction, however it is obviously not the most general possibility. In this section we will briefly consider more general couplings that are not symmetric. Now the cubic and quartic $AAA$ and $AAAA$ vertices (without $1/\Lambda^{2n}$ corrections) are not of the same form as Yang-Mills vertices. For example, the $\Lambda^0$ terms contributing to three point vertex $A_1A_2A_3$ are:
\be
\frac{1}{\sqrt{2}}f^{abc}\left(g_{123}\partial_{[\mu} A_{\nu]}^{1a}A^{2b\mu}A^{3c\nu}+g_{231}\partial_{[\mu} A_{\nu]}^{2b}A^{3c\mu}A^{1a\nu}+g_{312}\partial_{[\mu} A_{\nu]}^{3c}A^{1a\mu}A^{2b\nu}\right),
\ee
which gives a different Feynman rule than 3pt Yang-Mills vertex. For example, the on shell vertex coming from this term is 
\be
A_3(1^a,2^b,3^c)\propto  f_{abc}\left(-(g_{123}+g_{231})\epsilon_{1}\cdot\epsilon_{2}\;\epsilon_{3}\cdot p_{1}+(g_{123}+g_{312})\epsilon_{1}\cdot\epsilon_{3}\;\epsilon_{2}\cdot p_{1}-(g_{312}+g_{231})\epsilon_{1}\cdot p_{2}\;\epsilon_{2}\cdot\epsilon_{3}\right),
\ee
which is of different structure than Yang-Mills 3pt vertex,
\be
A_3(1^a,2^b,3^c)\propto f_{abc}\left(-\epsilon_{1}\cdot\epsilon_{2}\;\epsilon_{3}\cdot p_{1}+\epsilon_{1}\cdot\epsilon_{3}\;\epsilon_{2}\cdot p_{1}-\epsilon_{1}\cdot p_{2}\;\epsilon_{2}\cdot\epsilon_{3}\right).
\ee
Similarly we may consider non-symmetric $A_1A_2A_3A_4$ couplings:

\be
\begin{split} \label{eq:non sym AAAA}
\L_{AAAA}&=f^{abe}f^{cde}(g_{1234}A^{1a}\cdot A^{3c}A^{2b}\cdot A^{4d}+g_{1243}A^{1a}\cdot A^{4d}A^{2b}\cdot A^{3c})\\&+f^{ace}f^{bde}(g_{1324}A^{1a}\cdot A^{2b}A^{3c}\cdot A^{4d}+g_{1342}A^{1a}\cdot A^{4d}A^{3c}\cdot A^{2b})\\&+f^{ade}f^{bce}(g_{1423}A^{1a}\cdot A^{3c}A^{2b}\cdot A^{4d}+g_{1432}A^{1a}\cdot A^{2b}A^{3c}\cdot A^{4d}),
\end{split}
\ee

 \be
\begin{split}
\L_{AAAA}^{F^3}=&f^{abe}f^{cde}\bigg(G_{1234}  D_{[\mu}A_{1\nu]}^ a D^{[\nu}A_{2}^{b\rho]} A_{3\rho}^c A_4^{\mu d} +\\&G_{1324}  D_{[\mu}A_{1\nu]}^ a D^{[\nu}A_{3}^{b\rho]} A_{2\rho}^c A_4^{\mu d}+\\&G_{1423}  D_{[\mu}A_{1\nu]}^ a D^{[\nu}A_{4}^{b\rho]} A_{2\rho}^c A_3^{\mu d}+\\&G_{2413}  D_{[\mu}A_{2\nu]}^ a D^{[\nu}A_{4}^{b\rho]} A_{1\rho}^c A_3^{\mu d}+\\&G_{2314}  D_{[\mu}A_{2\nu]}^ a D^{[\nu}3_{4}^{b\rho]} A_{1\rho}^c A_4^{\mu d}+\\&G_{3412}  D_{[\mu}A_{3\nu]}^ a D^{[\nu}3_{4}^{b\rho]} A_{1\rho}^c A_2^{\mu d}\bigg),
\end{split}
\ee
where any other operators obtain by permuting ${1,2,3,4}$ labels of one of these operators does not give a new operator. Note that if we allow for non-symmetric couplings we do not need to write $\L_{AAAA2}^{F^3}$ term from \eqref{eq:HO} because these operators are already included in \eqref{eq:non sym AAAA}.
For simplicity we only considered $A_1A_2\rightarrow A_3 A_4$ scattering amplitude with these non-symmetric couplings up to $1/\Lambda^2$ order. We found that by imposing the BCJ relation we do not get a new solution, \ie all the couplings must be symmetric as before.

\section{Conclusion}\label{conclusion}
In this paper we explored the constraints that arise on a large class of low energy effective theories for a tower of interacting massive spin-1 states coupled to a massless gluon and scalar, by demanding that they respect colour-kinematic duality, a necessary precursor to a double copy, without introducing any spurious poles.
Since the kinematic factors that arise from Feynman diagram calculations do not automatically satisfy Jacobi identities, it is necessary to shift them in a manner which leaves the amplitude unchanged. As outlined in section~\ref{sec:spec}, when considering the scattering of massive states, in order to avoid spurious poles in the shifted kinematic factors, the spectrum of states must satisfy several `spectral conditions'. Once these are satisfied, the kinematic factors must further satisfy massive analogues of the BCJ relations. When all these conditions are met, the shifted kinematic factors respect colour-kinematic duality, and a local double copy should exist. \\

Since in general, the spectral conditions are difficult to solve, we have made the expedient choice that the spectrum of states should be identical to that that arises in Kaluza-Klein theories, specifically the standard compactification from five dimensions on an $S^1$, together with demanding the preservation of the associated global $U(1)$. This ensures that spectral conditions are identically satisfied. There is nevertheless a huge class of effective field theories which satisfy these requirements. Thus it is only necessary to impose the BCJ relations in order to ensure colour-kinematics duality is kept intact. Our analysis shows that that at least to quintic order in the Lagrangian, and up to order $1/\Lambda^4$ in the effective field theory expansion, the unique theory within our class which respects the colour-kinematics duality is the theory obtained from compactification of the 5d Lagrangian $\left(\frac{-1}{4}\text{tr}(F^2)+\frac{G_{5d}}{\Lambda^2}\text{tr}(F^3)-\frac{9G_{5d}^{2}}{16\Lambda^4}\text{tr}([F,F]^2)\right)$ on an $S^1$. The latter is of course known to admit a local double copy, including the higher derivative operators. Interestingly, while 4pt processes alone could not fix all of cubic and quartic couplings, we found that by combining them with 5pt processes we fixed all the cubic, quartic and quintic interactions. \\

Our results may be interpreted as the statement that Kaluza-Klein theory may be derived from the requirement that a specific spectrum of states which automatically satisfy the spectral conditions, admit a local double copy. While the fact that Kaluza-Klein theory is a consistent solution is itself not surprising, one might of thought that allowing for higher derivative/irrelevant operators in the EFT expansion would give us more freedom, and could be used to relax some of the constraints of the BCJ relations. To the order that we have calculated this situation does not arise. \\

It remains the case that there could still be some freedom in higher point amplitudes or at higher orders in the EFT expansion. At a given order in the EFT expansion, there are in general many other operators in the adjoint representation that could be included, in particular non-symmetric couplings of the type considered in section~\ref{nonsym}, as well as operators which do not appear in the Kaluza-Klein compactification of 5d Yang-Mills plus its higher derivative terms. Another possibility is to add multiple fields in different representations in 5d and to consider their compactification. We leave it to future works to explore or otherwise exclude these possibilities.\\

Given our results, it would be extremely helpful to have other nontrivial solutions of the spectral conditions. However, ultimately it may be necessary to relax the naive rules of the double copy to account for interacting theories with massive states, or similarly to account for higher order operators in an EFT expansion. By trying to mirror the massless double copy procedure as closely as possible, we have also forced ourselves into working with massive theories that are related to massless ones (in this case massless ones in higher dimensions), and to a large extent this explains the limitation of our results.

\bigskip
\noindent{\textbf{Acknowledgments:}}
 We would like to thank Mariana Carrillo Gonzalez for useful comments. The work of AJT is supported by STFC grants ST/P000762/1 and ST/T000791/1. JR is supported by an STFC studentship. AJT thanks the Royal Society for support at ICL through a Wolfson Research Merit Award.

\appendix

\section{BCJ relations}\label{appendixBCJ}

The $9 \times 9$ block diagonal matrix $A$ is as follows:
\begin{eqnarray}
\hspace{-15pt} A=\left(
\begin{array}{ccccccccc}
 B_1& D_{12}D_{34} & 0 & 0 & -D_{15}D_{34} & 0 &D_{25}D_{34} & 0 & 0 \\
 D_{12}D_{34} & B_2 & D_{12} D_{45} & 0 & 0 & 0 & 0 & 0 & 0 \\
 0 & D_{12} D_{45} &B_3 & D_{23} D_{45} & 0 & 0 & 0 &D_{13} D_{45} & 0 \\
 0 & 0 & D_{23} D_{45} & B_4 & D_{14} D_{23} & 0 & 0 & 0 \\
 -D_{15}D_{34} & 0 & 0 &D_{15} D_{23} &B_5& 0 & 0 & 0 & -D_{15} D_{24} \\
 0 & 0 & 0 & D_{14} D_{23} & 0 & B_6 & D_{14}D_{25} & 0 & 0 \\
D_{25}D_{34} & 0 & 0 & 0 & 0 & D_{14}D_{25} &B_7 & -D_{13}D_{25} & 0 \\
 0 & 0 &D_{13} D_{45} & 0 & 0 & 0 & -D_{13}D_{25} &B_8 &D_{13} D_{24} \\
 0 & 0 & 0 & 0 & -D_{15} D_{24} & 0 & 0 &D_{13} D_{24} &B_9 \\
\end{array}
\right),
\end{eqnarray}
where the $B_{i}$'s are:
\begin{equation}
    \begin{split}
        B_1=D_{12}D_{34}+D_{15} D_{34}+D_{25} D_{34},\\
        B_2=D_{12}D_{34}+D_{12} D_{35}+D_{12} D_{45},\\
        B_3=D_{12}D_{45}+D_{13} D_{45}+D_{23} D_{45},\\
        B_4=D_{14}D_{23}+D_{15} D_{23}+D_{23} D_{45},\\
        B_5=D_{15}D_{23}+D_{15} D_{24}+D_{15} D_{34},\\
        B_6=D_{14}D_{23}+D_{14} D_{25}+D_{14} D_{34},\\
        B_7=D_{13}D_{25}+D_{14} D_{25}+D_{25} D_{34},\\
        B_8=D_{13}D_{24}+D_{13} D_{25}+D_{13} D_{45},\\
        B_9=D_{13}D_{24}+D_{15} D_{24}+D_{24} D_{35}.\\
    \end{split}
\end{equation}
Imposing the spectral conditions \eqref{spectral2}, reduces the rank of $A$ from 9 to 5, where the following vectors form the basis for the null space of $A$:
\begin{equation}\label{nullspace}
    \begin{split}
    u_1^T&=\left(0,\frac{D_{24}}{D_{12}},-\frac{D_{24}}{D_{12}+D_{13}+D_{23}},0,-\frac{D_{24}}{D_{12}+D_{13}+D_{14}},0,0,0,1\right),\\
    u_2^T&=\left(-\frac{D_{13}}{D_{12}+D_{13}+D_{14}+D_{23}+D_{24}},0,-\frac{D_{13}}{D_{12}+D_{13}+D_{23}},0,-\frac{D_{13}}{D_{12}+D_{13}+D_{14}},0,0,1,0\right),\\
   u_3^T&= \left(-\frac{D_{12}+D_{23}+D_{24}}{D_{12}+D_{13}+D_{14}+D_{23}+D_{24}},0,-\frac{D_{12}+D_{23}+D_{24}}{D_{12}+D_{13}+D_{23}},-\frac{-D_{12}-D_{23}-D_{24}}{D_{23}},0,0,1,0,0\right),\\
    u_4^T&=\left(\frac{D_{14}}{D_{12}+D_{13}+D_{14}+D_{23}+D_{24}},\frac{D_{14}}{D_{12}},0,-\frac{D_{14}}{D_{23}},0,1,0,0,0\right).
    \end{split}
\end{equation}
These null vectors determine the BCJ relations via \eqref{BCJrelations}, expressed here as linear relations on the kinematic factors. To obtain \eqref{eq:bcj1},  \eqref{eq:bcj2},  \eqref{eq:bcj3} and \eqref{eq:bcj4} once the spectral condition is applied, we consider linear combinations of $U_\alpha^T $ in \eqref{BCJrelations},

\begin{equation}\label{linearcomb}
\sum_{\alpha} \beta_{\alpha}U_\alpha^T M n =0,
\end{equation}
the values of $\beta_{\alpha}$ for a given BCJ relation is given in the following table:
\begin{table}[H]
    \centering
    \begin{tabular}{|c|c|c|c|c|}
    \hline
         & $\beta_{1}$ & $\beta_{2}$ & $\beta_{3}$& $\beta_{4}$  \\
         \hline\hline
         \eqref{eq:bcj1} & $A_{1}$ & -$A_{1}$  &-$A_{1}$  &0 \\
         \hline
        \eqref{eq:bcj2} & -$A_{2}$   & $A_{2}$ &$A_{2}$ & $\frac{D_{24}}{D_{14}}A_{2}$ \\
         \hline
         \eqref{eq:bcj3} & 0 & -$A_{3}$ &-$A_{3}$&0 \\
         \hline
         \eqref{eq:bcj4} & 0 & $A_{4}$&$\frac{D_{12}+D_{23}}{D_{12}+D_{23}+D_{24}}A_{4}$ &0 \\
         \hline
    \end{tabular}
    \caption{Value of $\beta_{\alpha}$ in \eqref{linearcomb} in order to reproduce the four BCJ.}
    \label{linearcom}
\end{table}
where,
\begin{equation}
    \begin{split}
    A_1=&D_{12} (D_{12} + D_{13} + D_{14}) D_{23} (D_{12} + D_{13} + D_{14} + D_{23} + D_{24})\\
   A_2=&D_{14} (D_{12} + D_{13} + D_{14}) D_{23} (D_{12} + D_{13} + D_{14} + D_{23} + D_{24})\\
   A_3=&(D_{12} + D_{13} + D_{14}) D_{23} (D_{12} + D_{13} + D_{23}) (D_{12} + D_{13} + D_{14} + D_{23} + D_{24})\\
   A_4=&-(D_{12} + D_{13} + D_{14}) D_{23} (D_{12} + D_{23} + D_{24}) (D_{12} + D_{13} + D_{14} + D_{23} + D_{24})\\
    \end{split}
\end{equation}

\section{Interacting terms}\label{interactingterms}

 \begin{equation}
\begin{split}\label{eq:HO}
    &\L_{AAA}=\frac{1}{\sqrt{2}}f^{abc}\sum_{i,j,k \in \mathbb{Z}_{\ne 0}}\; g_{ijk}\left((D_{[\mu}A^{i}_{\nu]})^a A^{jb\mu}A^{kc\nu}\right),\\
    &\L_{AA\phi}=\frac{i}{\sqrt{2}}f^{abc}\sum_{i,j \in \mathbb{Z}_{\ne 0}}\; g'_{ij s}\left(A^{ia}_{\mu}A^{jb\mu}\phi^{ c}\right),\\
    &\L_{AAAA}=\frac{-1}{8}f^{abe}f^{cde}\sum_{i,j,k,l \in \mathbb{Z}_{\ne 0}}\; g_{ijkl}\left(A^{ia}_{[\mu}A^{jb}_{\nu]}A^{kc[\mu}A^{ld\nu]}\right),
    \\
    &\L_{AA\phi\phi}=\frac{-1}{4}f^{abe}f^{cde}\sum_{i,j \in \mathbb{Z}_{\ne 0}}\; g_{ijss}\left(A^{ia}_{\mu}A^{jc\mu}\phi^{b}\phi^{d}\right),\\
    &\L^{F^3}_{AAA1}=\frac{4}{\Lambda^2}f^{abc}\sum_{i,j,k \in \mathbb{Z}_{\ne 0}}\; G_{ijk}\left(D^{[\mu}A^{i\nu]}D_{\nu}A^{j\rho}D_{[\rho}A_{k\mu]}\right),\\
    &\L^{F^3}_{AAA2}=\frac{3}{\Lambda^2}f^{abc}\sum_{i,j,k \in \mathbb{Z}_{\ne 0}}\; \hat{G}_{ijk}\left(m_{i}m_{j}A^{ai\mu}A^{bj\nu}D_{[\nu}A^{kc}_{\mu]}\right),\\
    &\L^{F^3}_{AA\phi}=\frac{-6i}{\Lambda^2}f^{abc}\sum_{i,j \in \mathbb{Z}_{\ne 0}}\; G'_{ij s}\left(A^{ia\mu}D^{\rho}\phi^{b}D_{[\rho}A^{jc}_{\mu]}\right),\\
     &\L^{F^3}_{A\phi\phi}=\frac{3}{2\Lambda^2}f^{abc}  G_{0ss}\left(D^{\mu}\phi^{a}D^{\nu}\phi^{b}F_{\nu\mu}^{0c}\right),\\
    &\L^{F^3}_{AAAA1}=\frac{3\sqrt{2}}{\Lambda^2}f^{abe}f^{cde}\sum_{i,j,k \in \mathbb{Z}_{\ne 0}}\; G_{ijkl}\left(D^{[\mu}A^{ia\nu]}D_{[\nu}A^{jb}_{\rho]}A^{kc\rho}A^{dl}_{\mu}\right),
    \\
      &\L^{F^3}_{AAAA2}=\frac{-3}{2\sqrt{2}\Lambda^2}f^{abe}f^{cde}\sum_{i,j,k \in \mathbb{Z}_{\ne 0}}\; \hat{G}_{ijkl}\left(m_{i}m_{j}A^{ia}_{\mu}A^{jb\nu}A^{kc\mu}A^{ld}_{\nu}\right),
    \\
    &\L^{F^3}_{AA\phi\phi 1}=\frac{3}{2\sqrt{2}\Lambda^2}f^{abe}f^{cde}\sum_{i,j,k \in \mathbb{Z}_{\ne 0}}\; G_{ijss}\left(A^{ia}_{\mu}A^{jb}_{\nu}D^{\mu}\phi^{c}D^{\nu}\phi^{d}\right),
    \\
    &\L^{F^3}_{AA\phi\phi 2}=\frac{3\sqrt{2}}{\Lambda^2}f^{abe}f^{cde}\sum_{i,j,k \in \mathbb{Z}_{\ne 0}}\; \hat{G}_{ijss}\left(A^{ia\mu}\phi^{b}D_{[\mu}A^{jc}_{\nu]}D^{\nu}\phi^{d}\right),\\
    &\L^{F^3}_{AAA\phi 1}=\frac{-3i}{\sqrt{2}\Lambda^2}f^{abe}f^{cde}\sum_{i,j,k \in \mathbb{Z}_{\ne 0}}\; \hat{G}_{ijks}\left(m_kA^{ia}_{\mu}A^{jb}_{\nu} A^{kc\mu}D^{\nu}\phi^{d}\right),\\
    &\L^{F^3}_{AAA\phi 2}=\frac{-3\sqrt{2}i}{\Lambda^2}f^{abe}f^{cde}\sum_{i,j,k \in \mathbb{Z}_{\ne 0}}\; G_{ijks}\left(m_{i}A^{ja\mu}\phi^{b}A^{ic\nu}D_{[\mu}A^{kd}_{\nu]}\right),\\
     &\L^{F^3}_{AAAAA}=\frac{3}{2\Lambda^2}f^{a_1 b c}f^{b a_2 a_3}f^{c a_4 a_5}\sum_{i,j,k,l,m \in \mathbb{Z}_{\ne 0}}\; G_{ijklm}\left(D^{[\mu}A^{ia_1\nu]}A^{ja_2 \nu}A^{ka_3 \rho}A^{la_4}_{\rho}A^{ma_5 \mu}\right),\\
     &\L^{F^3}_{\phi AAAA}=\frac{3i}{2\Lambda^2}f^{a_1 b c}f^{b a_2 a_3}f^{c a_4 a_5}\sum_{i,j,k,l \in \mathbb{Z}_{\ne 0}}\; G_{ijkls}\left(m_i A^{ia_1\mu}\phi^{a_2} A^{ja_3\nu}A^{ka_4}_{\mu}A^{ma_5}_{\nu}\right),\\
     &\L^{F^3}_{\phi \phi AAA1}=\frac{-3}{2\Lambda^2}f^{a_1 b c}f^{b a_2 a_3}f^{c a_4 a_5}\sum_{i,j,k \in \mathbb{Z}_{\ne 0}}\; G_{ijkss}\left(D^{\mu}\phi^{a_1}\phi^{a_2} A^{ia_3\nu} A^{ja_4}_{\mu}A^{ka_5}_{\nu}\right),\\
     &\L^{F^3}_{\phi \phi AAA2}=\frac{3}{2\Lambda^2}f^{a_1 b c}f^{b a_2 a_3}f^{c a_4 a_5}\sum_{i,j,k \in \mathbb{Z}_{\ne 0}}\; \hat{G}_{ijks}\left(D_{[\mu} A^{ia_1}_{\nu]}\phi^{a_2} A^{ja_3\mu}\phi^{a_4} A^{ka_5\nu}\right),
\end{split}
\end{equation}

\begin{align}
\L^{F^4}_{AAAA1}&=\frac{-9}{\Lambda^4}f^{abe}f^{cde} \sum_{i,j,k,l \in \mathbb{Z}}\; c_{ijkl}\left(D^{[\mu}A^{ia\nu]}D_{[\alpha}A^{jb}_{\beta]}D_{\mu}A_{\nu}^{ck}D^{[\alpha}A^{ld\beta]}\right),
    \\ \nonumber
    \L^{F^4}_{AAAA2}&=\frac{9}{\Lambda^4}f^{abe}f^{cde} \sum_{i,j,k,l \in \mathbb{Z}}\; C_{ijkl}\left(m_iA^{ia\mu}D_{[\alpha}A^{jb}_{\beta]}m_kA_{\mu}^{ck}D^{[\alpha}A^{ld\beta]}\right),
    \\ \nonumber
    \L^{F^4}_{AAA\phi1}&=\frac{18i}{\Lambda^4}f^{abe}f^{cde} \sum_{i,j,k \in \mathbb{Z}}\; c_{ijks}\left(m_iA^{ia\mu}D_{[\alpha}A^{jb}_{\beta]}D_{\mu}\phi^{c}D^{[\alpha}A^{ld}_{\beta]}\right),
      \\ \nonumber
    \L^{F^4}_{AAA\phi2}&=\frac{-9i}{\Lambda^4}f^{abe}f^{cde} \sum_{i,j,k \in \mathbb{Z}}\; C_{ijks}\left(m_iA^{ia\mu}m_jA^{jb\nu}D_{\mu}\phi^{c}m_kA^{ld}_{\nu}\right),
       \\ \nonumber
    \L^{F^4}_{AA\phi\phi1}&=\frac{-9}{\Lambda^4}f^{abe}f^{cde} \sum_{i,j \in \mathbb{Z}}\; c_{ijss}\left(D^{[\mu}A^{ia\nu]}D_{\alpha}\phi^{b}D_{\mu}A_{\nu}^{ck}D^{\alpha}\phi^{d}\right),
    \\ \nonumber
    \L^{F^4}_{AA\phi\phi2}&=\frac{9}{2\Lambda^4}f^{abe}f^{cde} \sum_{i,j \in \mathbb{Z}}\; c^{(2)}_{ijss}\left(m_iA^{ia\mu}D_{\nu}\phi^{b}D_{\mu}\phi^{c}m_jA^{jd\nu}\right),\\ \nonumber
     \L^{F^4}_{AA\phi\phi3}&=\frac{9}{2\Lambda^4}f^{abe}f^{cde} \sum_{i,j \in \mathbb{Z}}\; c^{(3)}_{ijss}\left(m_iA^{ia\mu}D_{\nu}\phi^{b}D^{\nu}\phi^{c}m_jA_{jd\mu}\right),\\\nonumber
    \L^{F^4}_{\phi\phi\phi\phi}&=-\frac{9}{4\Lambda^4}f^{abe}f^{cde} \; c_{\phi4}\left(D^{\mu}\phi^{a}D_{\nu}\phi^{b}D_{\mu}\phi^{c}D^{\nu}\phi^{d}\right)\\\nonumber
     \L^{F^4}_{AAAAA1}&=\frac{-18}{\sqrt{2}\Lambda^4}f^{a_1 a_2 b}f^{a_3 c b}f^{c a_4 a_5}\sum_{i,j,k,l,m \in \mathbb{Z}_{\ne 0}}\; c_{ijklm}\left(D^{[\mu}A^{ia_1\nu]}D_{[\alpha}A^{ja_2}_{\beta]}D_{[\mu}A^{ka_3}_{\nu]}A^{la_4\alpha}A^{ma_5\beta}\right),\\\nonumber
     \L^{F^4}_{AAAAA2}&=\frac{9}{\sqrt{2}\Lambda^4}f^{a_1 a_2 b}f^{a_3 c b}f^{c a_4 a_5}\sum_{i,j,k,l,m \in \mathbb{Z}_{\ne 0}}\; C_{ijklm}\left(m_i m_k A^{ia_1\mu}D_{[\alpha}A^{ja_2}_{\beta]} A^{ka_3}_{\mu}A^{la_4\alpha}A^{ma_5\beta}\right),\\\nonumber    
     \L^{F^4}_{\phi AAAA1}&=\frac{-9i}{\sqrt{2}\Lambda^4}f^{a_1 a_2 b}f^{a_3 c b}f^{c a_4 a_5}\sum_{i,j,k,l \in \mathbb{Z}_{\ne 0}}\; c_{ijkls}\left( m_j D^{\mu}\phi^{a_1}D_{[\alpha}A^{ia_2}_{\beta]} A^{ja_3}_{\mu}A^{ka_4\alpha}A^{la_5\beta}\right),\\\nonumber \L^{F^4}_{\phi AAAA2}&=\frac{-18i}{\sqrt{2}\Lambda^4}f^{a_1 a_2 b}f^{a_3 c b}f^{c a_4 a_5}\sum_{i,j,k,l \in \mathbb{Z}_{\ne 0}}\; C_{ijkls}\left(m_j D^{[\mu}A^{ia_1\nu]}A^{ja_2}_{\beta}D_{[\mu}A^{ka_3}_{\nu]}\phi^{a_4}A^{la_5\beta}\right),\\  \nonumber    
     \L^{F^4}_{\phi AAAA3}&=\frac{-9i}{\sqrt{2}\Lambda^4}f^{a_1 a_2 b}f^{a_3 c b}f^{c a_4 a_5}\sum_{i,j,k,l \in \mathbb{Z}_{\ne 0}}\; \hat{C}_{ijkls}\left(m_i m_j m_k A^{ia_1\mu}A^{ja_2}_{\beta}A^{ka_3}_{\mu}\phi^{a_4}A^{la_5\beta}\right),\\\nonumber
     \L^{F^4}_{\phi \phi AAA1}&=\frac{9}{\sqrt{2}\Lambda^4}f^{a_1 a_2 b}f^{a_3 c b}f^{c a_4 a_5}\sum_{i,j,k \in \mathbb{Z}_{\ne 0}}\; c_{ijkss}\left( D^{\mu}\phi^{a_1}D_{[\alpha}A^{ia_2}_{\beta]} D_{\mu}\phi^{a_3}A^{ka_4\alpha}A^{la_5\beta}\right),\\\nonumber
     \L^{F^4}_{\phi \phi AAA2}&=\frac{18}{\sqrt{2}\Lambda^4}f^{a_1 a_2 b}f^{a_3 c b}f^{c a_4 a_5}\sum_{i,j,k \in \mathbb{Z}_{\ne 0}}\; C_{ijkss}\left(D^{[\mu}A^{ia_1\nu]}D_{\beta}\phi^{a_2} D_{[\mu}A^{ka_3}_{\nu]}\phi^{a_4}A^{la_5\beta}\right),\\\nonumber
     \L^{F^4}_{\phi \phi AAA3}&=\frac{-9}{\sqrt{2}\Lambda^4}f^{a_1 a_2 b}f^{a_3 c b}f^{c a_4 a_5}\sum_{i,j,k \in \mathbb{Z}_{\ne 0}}\; \hat{C}^{(3)}_{ijkss}\left(m_i m_j A^{ia_1\mu}D_{\beta}\phi^{a_2} A^{ja_3}_{\mu}\phi^{a_4}A^{ka_5\beta}\right),\\\nonumber
    \L^{F^4}_{\phi \phi AAA4}&=\frac{-9}{\sqrt{2}\Lambda^4}f^{a_1 a_2 b}f^{a_3 c b}f^{c a_4 a_5}\sum_{i,j,k \in \mathbb{Z}_{\ne 0}}\; \hat{C}^{(4)}_{ijkss}\left(m_i m_j A^{ia_1\mu}D_{\beta}\phi^{a_2} A^{ja_3\beta}\phi^{a_4}A^{ka_5}_{\mu}\right),\\\nonumber
    \L^{F^4}_{\phi \phi AAA5}&=\frac{-9}{\sqrt{2}\Lambda^4}f^{a_1 a_4 b}f^{ca_3 a_5}f^{a_2 b c}\sum_{i,j,k \in \mathbb{Z}_{\ne 0}}\; \hat{C}^{(5)}_{ijkss}\left(m_j m_k A^{ia_1\mu}D_{\beta}\phi^{a_2} A^{ja_3}_{\mu}\phi^{a_4}A^{ka_5\beta}\right),\\\nonumber
   \L^{F^4}_{\phi \phi \phi AA1}&=\frac{-9i}{\sqrt{2}\Lambda^4}f^{a_1 a_5 b}f^{a_2 bc}f^{a_3 a_4 c}\sum_{i,j \in \mathbb{Z}_{\ne 0}}\; c_{ijsss}\left(m_j A^{ia_1\mu}A^{ja_2\nu}D_{\mu}\phi^{a_3}D_{\nu}\phi^{a_4} \phi^{a_5}\right),\\\nonumber
      \L^{F^4}_{\phi \phi \phi AA2}&=\frac{-9i}{\sqrt{2}\Lambda^4}f^{a_1 a_4 b}f^{a_3 bc}f^{a_2 a_5 c}\sum_{i,j \in \mathbb{Z}_{\ne 0}}\; c^{(2)}_{ijsss}\left(m_i  A^{ia_1\mu}A^{ja_2\nu}D_{\mu}\phi^{a_3}D_{\nu}\phi^{a_4} \phi^{a_5}\right),\\\nonumber
     \L^{F^4}_{\phi \phi \phi AA3}&=\frac{-9}{\sqrt{2}\Lambda^4}f^{a_1 a_2 b}f^{a_3 c b}f^{c a_4 a_5}\sum_{i,j \in \mathbb{Z}_{\ne 0}}\; c^{(3)}_{ijsss}\left(m_i  D_{\mu}\phi^{a_1} A^{ia_2}_{\beta} D_{\mu}\phi^{a_3}\phi^{a_4}A^{la_5\beta}\right)   \, .
 \end{align}
Here $a,b,c,d,e$ are the indices of the adjoint representation of the gauge group and various numerical factors are chosen such that in the case of 4d KK reduction of 5d Yang-Mills with coupling $g$ plus $\frac{G}{\Lambda^2}\text{tr}(F^3)$ and $\frac{-9G^2}{16\Lambda^2}\text{tr}([F,F]^2)$ operators all couplings appear without any numerical factors, \ie they are expressed as products of $g$ and $G$.

\bibliographystyle{JHEP}
\bibliography{references}

\providecommand{\href}[2]{#2}\begingroup\raggedright\begin{thebibliography}{10}

\bibitem{Kawai:1985xq}
H.~Kawai, D.~C. Lewellen and S.~H.~H. Tye, \emph{{A Relation Between Tree
  Amplitudes of Closed and Open Strings}},
  \href{https://doi.org/10.1016/0550-3213(86)90362-7}{\emph{Nucl. Phys.}
  {\bfseries B269} (1986) 1}.

\bibitem{Bern:2010ue}
Z.~Bern, J.~J.~M. Carrasco and H.~Johansson, \emph{{Perturbative Quantum
  Gravity as a Double Copy of Gauge Theory}},
  \href{https://doi.org/10.1103/PhysRevLett.105.061602}{\emph{Phys. Rev. Lett.}
  {\bfseries 105} (2010) 061602}
  [\href{https://arxiv.org/abs/1004.0476}{{\ttfamily 1004.0476}}].

\bibitem{Cachazo:2014xea}
F.~Cachazo, S.~He and E.~Y. Yuan, \emph{{Scattering Equations and Matrices:
  From Einstein To Yang-Mills, DBI and NLSM}},
  \href{https://doi.org/10.1007/JHEP07(2015)149}{\emph{JHEP} {\bfseries 07}
  (2015) 149} [\href{https://arxiv.org/abs/1412.3479}{{\ttfamily 1412.3479}}].

\bibitem{Cheung:2017ems}
C.~Cheung, C.-H. Shen and C.~Wen, \emph{{Unifying Relations for Scattering
  Amplitudes}}, \href{https://doi.org/10.1007/JHEP02(2018)095}{\emph{JHEP}
  {\bfseries 02} (2018) 095}
  [\href{https://arxiv.org/abs/1705.03025}{{\ttfamily 1705.03025}}].

\bibitem{Chen:2013fya}
G.~Chen and Y.-J. Du, \emph{{Amplitude Relations in Non-linear Sigma Model}},
  \href{https://doi.org/10.1007/JHEP01(2014)061}{\emph{JHEP} {\bfseries 01}
  (2014) 061} [\href{https://arxiv.org/abs/1311.1133}{{\ttfamily 1311.1133}}].

\bibitem{Saotome:2012vy}
R.~Saotome and R.~Akhoury, \emph{{Relationship Between Gravity and Gauge
  Scattering in the High Energy Limit}},
  \href{https://doi.org/10.1007/JHEP01(2013)123}{\emph{JHEP} {\bfseries 01}
  (2013) 123} [\href{https://arxiv.org/abs/1210.8111}{{\ttfamily 1210.8111}}].

\bibitem{Monteiro:2014cda}
R.~Monteiro, D.~O'Connell and C.~D. White, \emph{{Black holes and the double
  copy}}, \href{https://doi.org/10.1007/JHEP12(2014)056}{\emph{JHEP} {\bfseries
  12} (2014) 056} [\href{https://arxiv.org/abs/1410.0239}{{\ttfamily
  1410.0239}}].

\bibitem{Luna:2015paa}
A.~Luna, R.~Monteiro, D.~O'Connell and C.~D. White, \emph{{The classical double
  copy for Taub–NUT spacetime}},
  \href{https://doi.org/10.1016/j.physletb.2015.09.021}{\emph{Phys. Lett.}
  {\bfseries B750} (2015) 272}
  [\href{https://arxiv.org/abs/1507.01869}{{\ttfamily 1507.01869}}].

\bibitem{Luna:2016due}
A.~Luna, R.~Monteiro, I.~Nicholson, D.~O'Connell and C.~D. White, \emph{{The
  double copy: Bremsstrahlung and accelerating black holes}},
  \href{https://doi.org/10.1007/JHEP06(2016)023}{\emph{JHEP} {\bfseries 06}
  (2016) 023} [\href{https://arxiv.org/abs/1603.05737}{{\ttfamily
  1603.05737}}].

\bibitem{White:2016jzc}
C.~D. White, \emph{{Exact solutions for the biadjoint scalar field}},
  \href{https://doi.org/10.1016/j.physletb.2016.10.052}{\emph{Phys. Lett.}
  {\bfseries B763} (2016) 365}
  [\href{https://arxiv.org/abs/1606.04724}{{\ttfamily 1606.04724}}].

\bibitem{Cardoso:2016amd}
G.~Cardoso, S.~Nagy and S.~Nampuri, \emph{{Multi-centered $ \mathcal{N}=2 $ BPS
  black holes: a double copy description}},
  \href{https://doi.org/10.1007/JHEP04(2017)037}{\emph{JHEP} {\bfseries 04}
  (2017) 037} [\href{https://arxiv.org/abs/1611.04409}{{\ttfamily
  1611.04409}}].

\bibitem{Luna:2016hge}
A.~Luna, R.~Monteiro, I.~Nicholson, A.~Ochirov, D.~O'Connell, N.~Westerberg
  et~al., \emph{{Perturbative spacetimes from Yang-Mills theory}},
  \href{https://doi.org/10.1007/JHEP04(2017)069}{\emph{JHEP} {\bfseries 04}
  (2017) 069} [\href{https://arxiv.org/abs/1611.07508}{{\ttfamily
  1611.07508}}].

\bibitem{Goldberger:2017frp}
W.~D. Goldberger, S.~G. Prabhu and J.~O. Thompson, \emph{{Classical gluon and
  graviton radiation from the bi-adjoint scalar double copy}},
  \href{https://doi.org/10.1103/PhysRevD.96.065009}{\emph{Phys. Rev.}
  {\bfseries D96} (2017) 065009}
  [\href{https://arxiv.org/abs/1705.09263}{{\ttfamily 1705.09263}}].

\bibitem{Ridgway:2015fdl}
A.~K. Ridgway and M.~B. Wise, \emph{{Static Spherically Symmetric Kerr-Schild
  Metrics and Implications for the Classical Double Copy}},
  \href{https://doi.org/10.1103/PhysRevD.94.044023}{\emph{Phys. Rev.}
  {\bfseries D94} (2016) 044023}
  [\href{https://arxiv.org/abs/1512.02243}{{\ttfamily 1512.02243}}].

\bibitem{De_Smet_2017}
P.-J. De~Smet and C.~D. White, \emph{Extended solutions for the biadjoint
  scalar field},
  \href{https://doi.org/10.1016/j.physletb.2017.11.007}{\emph{Physics Letters
  B} {\bfseries 775} (2017) 163–167}.

\bibitem{Bahjat_Abbas_2017}
N.~Bahjat-Abbas, A.~Luna and C.~D. White, \emph{The kerr-schild double copy in
  curved spacetime},
  \href{https://doi.org/10.1007/jhep12(2017)004}{\emph{Journal of High Energy
  Physics} {\bfseries 2017} (2017) }.

\bibitem{carrillogonzalez2017classical}
M.~Carrillo-Gonzalez, R.~Penco and M.~Trodden, \emph{The classical double copy
  in maximally symmetric spacetimes},
  \href{https://arxiv.org/abs/1711.01296}{{\ttfamily 1711.01296}}.

\bibitem{Goldberger_2018}
W.~D. Goldberger, J.~Li and S.~G. Prabhu, \emph{Spinning particles, axion
  radiation, and the classical double copy},
  \href{https://doi.org/10.1103/physrevd.97.105018}{\emph{Physical Review D}
  {\bfseries 97} (2018) }.

\bibitem{Li_2018}
J.~Li and S.~G. Prabhu, \emph{Gravitational radiation from the classical
  spinning double copy},
  \href{https://doi.org/10.1103/physrevd.97.105019}{\emph{Physical Review D}
  {\bfseries 97} (2018) }.

\bibitem{Lee_2018}
K.~Lee, \emph{Kerr-schild double field theory and classical double copy},
  \href{https://doi.org/10.1007/jhep10(2018)027}{\emph{Journal of High Energy
  Physics} {\bfseries 2018} (2018) }.

\bibitem{Plefka_2019}
J.~Plefka, J.~Steinhoff and W.~Wormsbecher, \emph{Effective action of dilaton
  gravity as the classical double copy of yang-mills theory},
  \href{https://doi.org/10.1103/physrevd.99.024021}{\emph{Physical Review D}
  {\bfseries 99} (2019) }.

\bibitem{Berman_2019}
D.~S. Berman, E.~Chacón, A.~Luna and C.~D. White, \emph{The self-dual
  classical double copy, and the eguchi-hanson instanton},
  \href{https://doi.org/10.1007/jhep01(2019)107}{\emph{Journal of High Energy
  Physics} {\bfseries 2019} (2019) }.

\bibitem{Kim:2019jwm}
K.~Kim, K.~Lee, R.~Monteiro, I.~Nicholson and D.~Peinador~Veiga, \emph{{The
  Classical Double Copy of a Point Charge}},
  \href{https://doi.org/10.1007/JHEP02(2020)046}{\emph{JHEP} {\bfseries 02}
  (2020) 046} [\href{https://arxiv.org/abs/1912.02177}{{\ttfamily
  1912.02177}}].

\bibitem{Goldberger:2019xef}
W.~D. Goldberger and J.~Li, \emph{{Strings, extended objects, and the classical
  double copy}}, \href{https://doi.org/10.1007/JHEP02(2020)092}{\emph{JHEP}
  {\bfseries 02} (2020) 092}
  [\href{https://arxiv.org/abs/1912.01650}{{\ttfamily 1912.01650}}].

\bibitem{Alawadhi:2019urr}
R.~Alawadhi, D.~Peinador~Veiga, D.~S. Berman and B.~Spence, \emph{{S-duality
  and the double copy}},
  \href{https://doi.org/10.1007/JHEP03(2020)059}{\emph{JHEP} {\bfseries 03}
  (2020) 059} [\href{https://arxiv.org/abs/1911.06797}{{\ttfamily
  1911.06797}}].

\bibitem{Banerjee:2019saj}
A.~Banerjee, E.~Colg\'ain, J.~Rosabal and H.~Yavartanoo, \emph{{Ehlers as EM
  duality in the double copy}},
  \href{https://arxiv.org/abs/1912.02597}{{\ttfamily 1912.02597}}.

\bibitem{Huang:2019cja}
Y.-T. Huang, U.~Kol and D.~O'Connell, \emph{{The Double Copy of
  Electric-Magnetic Duality}},
  \href{https://arxiv.org/abs/1911.06318}{{\ttfamily 1911.06318}}.

\bibitem{Berman:2020xvs}
D.~S. Berman, K.~Kim and K.~Lee, \emph{{The Classical Double Copy for M-theory
  from a Kerr-Schild Ansatz for Exceptional Field Theory}},
  \href{https://arxiv.org/abs/2010.08255}{{\ttfamily 2010.08255}}.

\bibitem{Neill:2013wsa}
D.~Neill and I.~Z. Rothstein, \emph{{Classical Space-Times from the S Matrix}},
  \href{https://doi.org/10.1016/j.nuclphysb.2013.09.007}{\emph{Nucl. Phys. B}
  {\bfseries 877} (2013) 177}
  [\href{https://arxiv.org/abs/1304.7263}{{\ttfamily 1304.7263}}].

\bibitem{Bern:2019crd}
Z.~Bern, C.~Cheung, R.~Roiban, C.-H. Shen, M.~P. Solon and M.~Zeng,
  \emph{{Black Hole Binary Dynamics from the Double Copy and Effective
  Theory}}, \href{https://doi.org/10.1007/JHEP10(2019)206}{\emph{JHEP}
  {\bfseries 10} (2019) 206}
  [\href{https://arxiv.org/abs/1908.01493}{{\ttfamily 1908.01493}}].

\bibitem{Shen:2018ebu}
C.-H. Shen, \emph{{Gravitational Radiation from Color-Kinematics Duality}},
  \href{https://doi.org/10.1007/JHEP11(2018)162}{\emph{JHEP} {\bfseries 11}
  (2018) 162} [\href{https://arxiv.org/abs/1806.07388}{{\ttfamily
  1806.07388}}].

\bibitem{Bautista:2019tdr}
Y.~F. Bautista and A.~Guevara, \emph{{From Scattering Amplitudes to Classical
  Physics: Universality, Double Copy and Soft Theorems}},
  \href{https://arxiv.org/abs/1903.12419}{{\ttfamily 1903.12419}}.

\bibitem{Cheung:2020gyp}
C.~Cheung and M.~P. Solon, \emph{{Classical gravitational scattering at $
  \mathcal{O} $(G$^{3}$) from Feynman diagrams}},
  \href{https://doi.org/10.1007/JHEP06(2020)144}{\emph{JHEP} {\bfseries 06}
  (2020) 144} [\href{https://arxiv.org/abs/2003.08351}{{\ttfamily
  2003.08351}}].

\bibitem{Carrillo-Gonzalez:2019aao}
M.~Carrillo~González, R.~Penco and M.~Trodden, \emph{{Shift symmetries, soft
  limits, and the double copy beyond leading order}},
  \href{https://arxiv.org/abs/1908.07531}{{\ttfamily 1908.07531}}.

\bibitem{Low:2019wuv}
I.~Low and Z.~Yin, \emph{{New Flavor-Kinematics Dualities and Extensions of
  Nonlinear Sigma Models}},  \href{https://arxiv.org/abs/1911.08490}{{\ttfamily
  1911.08490}}.

\bibitem{Carrasco:2019qwr}
J.~J.~M. Carrasco and L.~Rodina, \emph{{UV considerations on scattering
  amplitudes in a web of theories}},
  \href{https://doi.org/10.1103/PhysRevD.100.125007}{\emph{Phys. Rev.}
  {\bfseries D100} (2019) 125007}
  [\href{https://arxiv.org/abs/1908.08033}{{\ttfamily 1908.08033}}].

\bibitem{Carrasco:2019yyn}
J.~J.~M. Carrasco, L.~Rodina, Z.~Yin and S.~Zekioglu, \emph{{Simple encoding of
  higher derivative gauge and gravity counterterms}},
  \href{https://arxiv.org/abs/1910.12850}{{\ttfamily 1910.12850}}.

\bibitem{Low:2020ubn}
I.~Low, L.~Rodina and Z.~Yin, \emph{{Double Copy in Higher Derivative Operators
  of Nambu-Goldstone Bosons}},
  \href{https://arxiv.org/abs/2009.00008}{{\ttfamily 2009.00008}}.

\bibitem{Cheung:2020qxc}
C.~Cheung, J.~Mangan and C.-H. Shen, \emph{{Hidden Conformal Invariance of
  Scalar Effective Field Theories}},
  \href{https://arxiv.org/abs/2005.13027}{{\ttfamily 2005.13027}}.

\bibitem{Rodina:2020jlw}
L.~Rodina, \emph{{UV consistency conditions for Cachazo-He-Yuan integrands}},
  \href{https://doi.org/10.1103/PhysRevD.102.045012}{\emph{Phys. Rev. D}
  {\bfseries 102} (2020) 045012}
  [\href{https://arxiv.org/abs/2005.06446}{{\ttfamily 2005.06446}}].

\bibitem{Bern:2008qj}
Z.~Bern, J.~J.~M. Carrasco and H.~Johansson, \emph{{New Relations for
  Gauge-Theory Amplitudes}},
  \href{https://doi.org/10.1103/PhysRevD.78.085011}{\emph{Phys. Rev.}
  {\bfseries D78} (2008) 085011}
  [\href{https://arxiv.org/abs/0805.3993}{{\ttfamily 0805.3993}}].

\bibitem{Naculich:2014naa}
S.~G. Naculich, \emph{{Scattering equations and BCJ relations for gauge and
  gravitational amplitudes with massive scalar particles}},
  \href{https://doi.org/10.1007/JHEP09(2014)029}{\emph{JHEP} {\bfseries 09}
  (2014) 029} [\href{https://arxiv.org/abs/1407.7836}{{\ttfamily 1407.7836}}].

\bibitem{Chiodaroli:2015rdg}
M.~Chiodaroli, M.~Gunaydin, H.~Johansson and R.~Roiban, \emph{{Spontaneously
  Broken Yang-Mills-Einstein Supergravities as Double Copies}},
  \href{https://doi.org/10.1007/JHEP06(2017)064}{\emph{JHEP} {\bfseries 06}
  (2017) 064} [\href{https://arxiv.org/abs/1511.01740}{{\ttfamily
  1511.01740}}].

\bibitem{Johansson:2015oia}
H.~Johansson and A.~Ochirov, \emph{{Color-Kinematics Duality for QCD
  Amplitudes}}, \href{https://doi.org/10.1007/JHEP01(2016)170}{\emph{JHEP}
  {\bfseries 01} (2016) 170}
  [\href{https://arxiv.org/abs/1507.00332}{{\ttfamily 1507.00332}}].

\bibitem{Chiodaroli:2017ehv}
M.~Chiodaroli, M.~Gunaydin, H.~Johansson and R.~Roiban, \emph{{Gauged
  Supergravities and Spontaneous Supersymmetry Breaking from the Double Copy
  Construction}},
  \href{https://doi.org/10.1103/PhysRevLett.120.171601}{\emph{Phys. Rev. Lett.}
  {\bfseries 120} (2018) 171601}
  [\href{https://arxiv.org/abs/1710.08796}{{\ttfamily 1710.08796}}].

\bibitem{Chiodaroli:2018dbu}
M.~Chiodaroli, M.~Günaydin, H.~Johansson and R.~Roiban, \emph{{Non-Abelian
  gauged supergravities as double copies}},
  \href{https://doi.org/10.1007/JHEP06(2019)099}{\emph{JHEP} {\bfseries 06}
  (2019) 099} [\href{https://arxiv.org/abs/1812.10434}{{\ttfamily
  1812.10434}}].

\bibitem{bautista2019double}
Y.~F. Bautista and A.~Guevara, \emph{On the double copy for spinning matter},
  \href{https://arxiv.org/abs/1908.11349}{{\ttfamily 1908.11349}}.

\bibitem{Bjerrum-Bohr:2019nws}
N.~Bjerrum-Bohr, A.~Cristofoli, P.~H. Damgaard and H.~Gomez,
  \emph{{Scalar-Graviton Amplitudes}},
  \href{https://doi.org/10.1007/JHEP11(2019)148}{\emph{JHEP} {\bfseries 11}
  (2019) 148} [\href{https://arxiv.org/abs/1908.09755}{{\ttfamily
  1908.09755}}].

\bibitem{Johansson:2019dnu}
H.~Johansson and A.~Ochirov, \emph{{Double copy for massive quantum particles
  with spin}}, \href{https://doi.org/10.1007/JHEP09(2019)040}{\emph{JHEP}
  {\bfseries 09} (2019) 040}
  [\href{https://arxiv.org/abs/1906.12292}{{\ttfamily 1906.12292}}].

\bibitem{Plefka:2019wyg}
J.~Plefka, C.~Shi and T.~Wang, \emph{{The Double Copy of Massive Scalar-QCD}},
  \href{https://doi.org/10.1103/PhysRevD.101.066004}{\emph{Phys. Rev.}
  {\bfseries D101} (2020) 066004}
  [\href{https://arxiv.org/abs/1911.06785}{{\ttfamily 1911.06785}}].

\bibitem{bautista2019scattering}
Y.~F. Bautista and A.~Guevara, \emph{From scattering amplitudes to classical
  physics: Universality, double copy and soft theorems},
  \href{https://arxiv.org/abs/1903.12419}{{\ttfamily 1903.12419}}.

\bibitem{Carrasco:2020ywq}
J.~J.~M. Carrasco and I.~A. Vazquez-Holm, \emph{{Loop-Level Double-Copy for
  Massive Quantum Particles}},
  \href{https://arxiv.org/abs/2010.13435}{{\ttfamily 2010.13435}}.

\bibitem{Moynihan:2020ejh}
N.~Moynihan, \emph{{Scattering Amplitudes and the Double Copy in Topologically
  Massive Theories}},  \href{https://arxiv.org/abs/2006.15957}{{\ttfamily
  2006.15957}}.

\bibitem{Momeni:2020vvr}
A.~Momeni, J.~Rumbutis and A.~J. Tolley, \emph{{Massive Gravity from Double
  Copy}}, \href{https://doi.org/10.1007/JHEP12(2020)030}{\emph{JHEP} {\bfseries
  12} (2020) 030} [\href{https://arxiv.org/abs/2004.07853}{{\ttfamily
  2004.07853}}].

\bibitem{Johnson:2020pny}
L.~A. Johnson, C.~R.~T. Jones and S.~Paranjape, \emph{Constraints on a massive
  double-copy and applications to massive gravity},
  \href{https://arxiv.org/abs/2004.12948}{{\ttfamily 2004.12948}}.

\bibitem{deRham:2010ik}
C.~de~Rham and G.~Gabadadze, \emph{{Generalization of the Fierz-Pauli Action}},
  \href{https://doi.org/10.1103/PhysRevD.82.044020}{\emph{Phys. Rev.}
  {\bfseries D82} (2010) 044020}
  [\href{https://arxiv.org/abs/1007.0443}{{\ttfamily 1007.0443}}].

\bibitem{deRham:2010kj}
C.~de~Rham, G.~Gabadadze and A.~J. Tolley, \emph{{Resummation of Massive
  Gravity}}, \href{https://doi.org/10.1103/PhysRevLett.106.231101}{\emph{Phys.
  Rev. Lett.} {\bfseries 106} (2011) 231101}
  [\href{https://arxiv.org/abs/1011.1232}{{\ttfamily 1011.1232}}].

\bibitem{Hinterbichler:2015pqa}
K.~Hinterbichler and A.~Joyce, \emph{{Hidden symmetry of the Galileon}},
  \href{https://doi.org/10.1103/PhysRevD.92.023503}{\emph{Phys. Rev.}
  {\bfseries D92} (2015) 023503}
  [\href{https://arxiv.org/abs/1501.07600}{{\ttfamily 1501.07600}}].

\bibitem{deRham:2017zjm}
C.~de~Rham, S.~Melville, A.~J. Tolley and S.-Y. Zhou, \emph{{UV complete me:
  Positivity Bounds for Particles with Spin}},
  \href{https://doi.org/10.1007/JHEP03(2018)011}{\emph{JHEP} {\bfseries 03}
  (2018) 011} [\href{https://arxiv.org/abs/1706.02712}{{\ttfamily
  1706.02712}}].

\bibitem{deRham:2018qqo}
C.~de~Rham, S.~Melville, A.~J. Tolley and S.-Y. Zhou, \emph{{Positivity Bounds
  for Massive Spin-1 and Spin-2 Fields}},
  \href{https://arxiv.org/abs/1804.10624}{{\ttfamily 1804.10624}}.

\bibitem{Feng:2010my}
B.~Feng, R.~Huang and Y.~Jia, \emph{{Gauge Amplitude Identities by On-shell
  Recursion Relation in S-matrix Program}},
  \href{https://doi.org/10.1016/j.physletb.2010.11.011}{\emph{Phys. Lett. B}
  {\bfseries 695} (2011) 350}
  [\href{https://arxiv.org/abs/1004.3417}{{\ttfamily 1004.3417}}].

\bibitem{Jia:2010nz}
Y.~Jia, R.~Huang and C.-Y. Liu, \emph{{$U(1)$-decoupling, KK and BCJ relations
  in $\mathcal{N}=4$ SYM}},
  \href{https://doi.org/10.1103/PhysRevD.82.065001}{\emph{Phys. Rev. D}
  {\bfseries 82} (2010) 065001}
  [\href{https://arxiv.org/abs/1005.1821}{{\ttfamily 1005.1821}}].

\bibitem{Chen:2011jxa}
Y.-X. Chen, Y.-J. Du and B.~Feng, \emph{{A Proof of the Explicit Minimal-basis
  Expansion of Tree Amplitudes in Gauge Field Theory}},
  \href{https://doi.org/10.1007/JHEP02(2011)112}{\emph{JHEP} {\bfseries 02}
  (2011) 112} [\href{https://arxiv.org/abs/1101.0009}{{\ttfamily 1101.0009}}].

\bibitem{Bonifacio:2019ioc}
J.~Bonifacio and K.~Hinterbichler, \emph{{Unitarization from Geometry}},
  \href{https://doi.org/10.1007/JHEP12(2019)165}{\emph{JHEP} {\bfseries 12}
  (2019) 165} [\href{https://arxiv.org/abs/1910.04767}{{\ttfamily
  1910.04767}}].

\bibitem{broedel2012color}
J.~Broedel and L.~J. Dixon, \emph{{Color-kinematics duality and double-copy
  construction for amplitudes from higher-dimension operators}},
  \href{https://doi.org/10.1007/JHEP10(2012)091}{\emph{JHEP} {\bfseries 10}
  (2012) 091} [\href{https://arxiv.org/abs/1208.0876}{{\ttfamily 1208.0876}}].

\end{thebibliography}\endgroup

\end{document}